\documentclass[a4paper,onecolumn,11pt,accepted=2023-10-04]{quantumarticle}
\pdfoutput=1
\usepackage[utf8]{inputenc}
\usepackage[english]{babel}
\usepackage[T1]{fontenc}
\usepackage{amsmath}
\usepackage{amsthm}
\usepackage{hyperref}
\usepackage{cleveref}

\usepackage{tikz}
\usepackage{lipsum}

\newcommand{\mc}{\mathcal}

\newcommand{\ket}[1]{\ensuremath{\lvert #1 \rangle}}

\newcommand{\tr}[1]{\text{tr}(#1)}

\begin{document}

\title{Synergetic quantum error mitigation by randomized compiling and zero-noise extrapolation for the variational quantum eigensolver}
\author{Tomochika Kurita}
\email{kurita.tomo@fujitsu.com}
\orcid{0000-0002-2445-2701}
\affiliation{Quantum Laboratory, Fujitsu Research, Fujitsu Limited. 10-1 Morinosato-wakamiya, Atsugi, Kanagawa, Japan 243-0197}
\author{Hammam Qassim}
\affiliation{Keysight Technologies Canada, 137 Glasgow St, Kitchener, ON, Canada, N2G 4X8}
\author{Masatoshi Ishii}
\affiliation{Quantum Laboratory, Fujitsu Research, Fujitsu Limited. 10-1 Morinosato-wakamiya, Atsugi, Kanagawa, Japan 243-0197}
\author{Hirotaka Oshima}
\affiliation{Quantum Laboratory, Fujitsu Research, Fujitsu Limited. 10-1 Morinosato-wakamiya, Atsugi, Kanagawa, Japan 243-0197}
\author{Shintaro Sato}
\affiliation{Quantum Laboratory, Fujitsu Research, Fujitsu Limited. 10-1 Morinosato-wakamiya, Atsugi, Kanagawa, Japan 243-0197}
\author{Joseph Emerson}
\affiliation{Keysight Technologies Canada, 137 Glasgow St, Kitchener, ON, Canada, N2G 4X8}

\maketitle

\begin{abstract}
We propose a quantum error mitigation strategy for the variational quantum eigensolver (VQE) algorithm. We find, via numerical simulation, that very small amounts of coherent noise in VQE can cause substantially large errors that are difficult to suppress by conventional mitigation methods, and yet our proposed mitigation strategy is able to significantly reduce these errors. The proposed strategy is a combination of previously reported techniques, namely randomized compiling (RC) and zero-noise extrapolation (ZNE). Intuitively, randomized compiling turns coherent errors in the circuit into stochastic Pauli errors, which facilitates extrapolation to the zero-noise limit when evaluating the cost function. Our numerical simulation of VQE for small molecules shows that the proposed strategy can mitigate energy errors induced by various types of coherent noise by up to two orders of magnitude.
\end{abstract}

\section{Introduction}\label{sec intro}
Quantum computing is an emerging technology with the potential to solve classically intractable computational problems. One of its most promising applications is simulation of quantum chemistry \cite{mcardle2020quantum,paudel2022quantum,rice2021quantum}. Standard approaches for solving quantum chemistry problems on a quantum computer require a fault-tolerant implementation \cite{fowler2012surface}, which unfortunately remains a distant future prospect. The variational quantum eigensolver (VQE) algorithm was proposed as a near-term alternative \cite{peruzzo2014variational,mcclean2016theory}. 
VQE is based on the hybrid quantum-classical computational paradigm, where a classical computer guides a parameterized quantum circuit towards the correct solution. In recent years, proof-of-principle demonstrations of VQE were successfully conducted on current, noisy intermediate-scale quantum (NISQ) computers \cite{rice2021quantum, peruzzo2014variational, o2016scalable, kandala2017hardware, colless2018computation, kandala2019error, shen2017quantum, nam2020ground}. 

Scaling VQE to larger system sizes seems to pose serious challenges. For example it is known that variational algorithms suffer from the barren plateau phenomenon, where the energy landscape is exponentially flat in most of the Hilbert space \cite{mcclean2018barren}. Additionally, when the circuit ansatz contains too many parameters the classical optimization step could become too slow to converge \cite{tilly2022variational}. These problems are greatly exacerbated in the presence of noise. Studying the complex interplay in VQE between noise, classical optimization, and circuit ansatz is a major motivation of this work. As the error tolerance required for useful chemistry calculations is very small (< $1.6$ mHa), it is highly desirable to find ways to suppress and mitigate the effect of different noise processes in VQE.

To address the problem of noise on near-term devices, several quantum error mitigation schemes have been proposed to suppress the effect of quantum noise and correct the expectation values of observables of interest \cite{endo2021hybrid}. Unlike quantum error correction, error mitigation schemes do not require additional qubits, but generally require additional circuit runs and possibly some increase in circuit depth. Most critically, the accuracy of these methods depends on very strong, physically unmotivated or difficult-to-verify assumptions about the nature of the error model affecting the computation. Overcoming the latter limitation is another major motivation of the current work. 
 
A few variants of quantum error mitigation are known, some of which specifically target VQE applications. For example, zero-noise extrapolation (ZNE) \cite{kandala2019error, li2017efficient, temme2017error, he2020zero} and probabilistic error cancellation (PEC) \cite{temme2017error, he2020zero, zhang2020error} are effective for mitigating Pauli-stochastic noise. The quantum subspace expansion (QSE) \cite{colless2018computation, mcclean2017hybrid} was suggested to mitigate errors in VQE, and was argued to be especially effective for mitigating the effect of decoherence. However, these methods are not effective in the presence of coherent noise, and while some kinds of coherent noise can be reduced by better calibration of the hardware, other kinds, such as coherent crosstalk, pose significant challenges with current calibration technology. Furthermore, even a small miscalibration can result in significant error in the algorithm's output if the computation is relatively long. 
Randomized compiling (RC) \cite{wallman2016noise} is a compilation technique with an important noise-tailoring effect, in which coherent noise processes in a noisy circuit are transformed into effective incoherent noise processes, with a marked reduction in the worst-case error rate. The benefit of RC for practical algorithms has been demonstrated in experiment \cite{hashim2021randomized}, as well as methods that combine RC with quantum error mitigation protocols \cite{ville2021leveraging, kim2021scalable, song2019quantum, ware2021experimental, ferracin2022efficiently, blunt2023statistical}. To the best of our knowledge the effect of combining RC with quantum error mitigation on the accuracy of a VQE algorithm has not been reported thus far. 

In this work we numerically simulate noisy VQE experiments and show significant improvement in their performance under an error mitigation strategy that combines RC and ZNE.
We give evidence that this improvement is generic by performing simulations under varied conditions, such as different target molecules, noise models, and classical optimizers.
We use a structured variational ansatz, namely the unitary coupled cluster ansatz with single and double excitations (UCC-SD) \cite{mcclean2016theory}. The main advantage from using a structured ansatz is that it explores a small subset of the Hilbert space, in which there is good reason to believe the ground state lies -- thus it requires fewer circuit parameters and is less likely to exhibit a vanishing gradient in the error-free case \cite{wang2021noise}.

In our simulations we use an error model in which single-qubit gates are perfect and two-qubit gates are noisy, where the noise is described by a small unitary error, possibly entangling the target qubits with other qubits. In particular, decoherence and relaxation noise is assumed to be negligible. This represents a situation where the circuit can be executed without loss of coherence, but still suffers from systematic noise due to small calibration errors or small amounts of crosstalk.
Such errors can quickly add up in a long circuit and result in a poor output state, greatly harming the performance of VQE. In addition, coherent noise of this form can be harmful if the variational ansatz is highly specialized to explore a small subset of the full Hilbert space. In this case coherent noise can take the state outside of the relevant manifold, resulting in unphysical results.

Our findings suggest that the application of both RC and ZNE to deep, structured VQE algorithms can provide a drastic synergetic effect in terms of the accuracy of the molecular energy extracted from VQE under these conditions. Interestingly, under all coherent noise models we consider, we find that the effects of RC and ZNE when applied separately are rather limited, but combining them reduces the energy error by up to two orders of magnitude.

The circuits we simulate are in a regime where the depth is challenging for NISQ devices, but much smaller than the depth required for non-variational methods such as quantum phase estimation (QPE) \cite{nielsen2002quantum}.
There are two motivations for studying this regime. First, deep VQE algorithms may become practical in a post NISQ era where the error rates are much smaller than in NISQ devices, but still too high for QPE. Second, QPE requires as input a good approximation to the ground state, and it has been recently argued that this requirement severely limits the usefulness of QPE for chemistry simulations \cite{lee2022there}. Here we note one strategy that could circumvent the aforementioned argument: when successful, a quantum variational algorithm (such as VQE) could find a circuit that efficiently prepares a state with high overlap with the ground state, which can then be used as input to QPE.  

The outline of the paper is as follows. In \Cref{sec vqe}, we explain background theory of VQE and the error-mitigation method employed in this work, as well as our noise models and simulation methodology. In \Cref{sec: results}, we present and discuss our simulation results. In \Cref{sec conclusion}, we conclude and discuss future perspectives.

\section{Variational quantum eigensolver simulation and noise models}\label{sec vqe}

\subsection{Variational quantum eigensolver}\label{subsec: vqe}
\begin{figure*}[t]
    \centering
    \includegraphics[scale = 1]{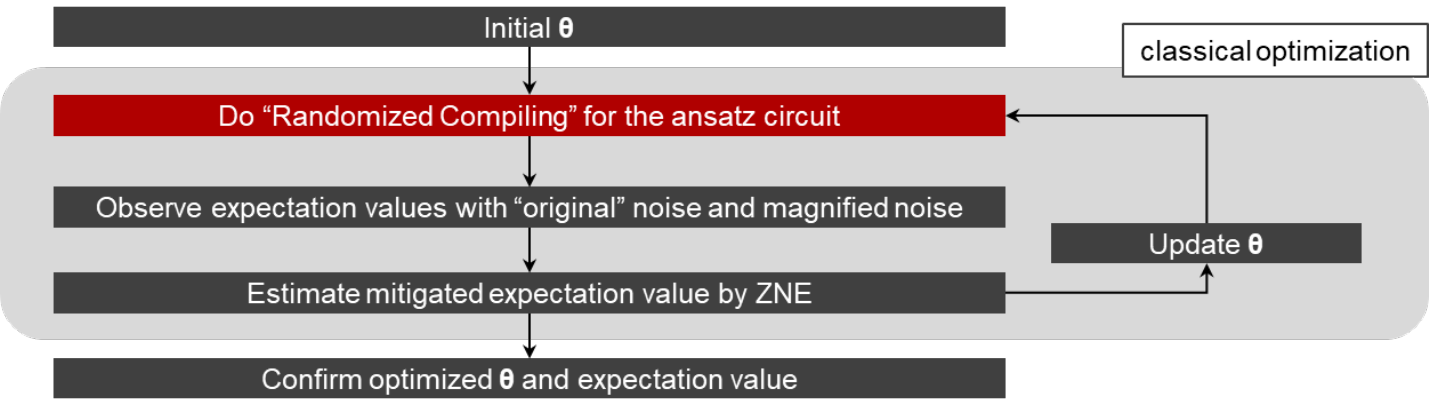}
    \vspace*{-3mm}
    \caption{Overall VQE scheme with RC and ZNE}
    \label{fig:overall_scheme}
\end{figure*}%
VQE is a hybrid classical-quantum algorithm for simulating properties of a Hamiltonian $\mc{H}$. 
In this work $\mathcal{H}$ represents the electronic structure Hamiltonian of some small molecule. 
VQE is based on using a classical optimizer to minimize the expected energy over some parameterized family of quantum circuits, where the cost function is evaluated within the optimization loop using a quantum computer. When successful, this algorithm finds a circuit that prepares an approximation to the ground state of the molecule, as well as an estimate of its ground state energy. 

In this work we study a family of VQE experiments designed to compute the dependence of the ground state energy on the positions of the nuclei in the molecule, reproducing the so-called potential energy surface of the molecule (or a one-dimensional slice of it, commonly known as a disassociation curve). To embed $\mathcal{H}$ into qubits, we use the standard techniques of second quantization and the Jordan-Wigner transform.
For a variational ansatz, we use the unitary coupled cluster ansatz with single- and double-excitations (UCC-SD) . The parameterized states have the form
\begin{align}
|\psi(\theta) \rangle = U(\theta) | \psi_{\text{HF}} \rangle
\end{align}
where $\theta = \theta_1, \dots, \theta_m$ are the variational parameters, and $| \psi_{\text{HF}} \rangle$ is a computational basis state encoding the Hartree-Fock approximation to the ground state. The circuit $U(\theta)$ is a Trotterized version of the exponential $e^{T(\theta) - T(\theta)^\dagger}$ where $T(\theta)$ is the cluster operator with parameters $\theta$. For a given value $\theta$, the cost function is the expected energy, $\langle \psi(\theta) |\mathcal{H}| \psi(\theta) \rangle$, which can be computed efficiently on the quantum computer. See \Cref{appdx: vqe} for more details.

\subsection{Noise models}
We choose a noise model in which entangling gates are noisy and single-qubit gates are perfect. The motivation is that we are mainly concerned with simulating residual calibration errors, and single-qubit gates can typically be calibrated to orders-of-magnitude higher fidelity than two-qubit gates. We choose an error rate of roughly $10^{-3}$. In comparison the circuits we simulate have up to $10^3$ noisy two-qubit gates.
In our work $U(\theta)$ is compiled using a gateset consisting of $\mathtt{CNOT}$ gates and arbitrary single-qubit gates. We consider four kinds of coherent and time stationary noise models affecting the $\mathtt{CNOT}$: (i) over-rotation by a small amount $\epsilon$ (that is $U \mapsto UU^\epsilon$), (ii) small X-axis rotation on the control qubit prior to the $\mathtt{CNOT}$ gate (we call it XI rotation noise hereafter), (iii) small Z-axis rotation on the target qubit prior to the $\mathtt{CNOT}$ gate (we call it IZ rotation noise hereafter), and (iv) coherent crosstalk where a small unwanted ZZ interaction couples each of the two qubits involved in a $\mathtt{CNOT}$ gate and its nearest neighbours, where the qubits are assumed to be arranged in a 1D chain with periodic boundaries. 
Other sources of noise, such as relaxation processes and state-preparation and measurement (SPAM) errors are not considered in this work.  

\subsection{Simulation details}

\begin{figure*}
    \centering
    \includegraphics[scale = 0.208]{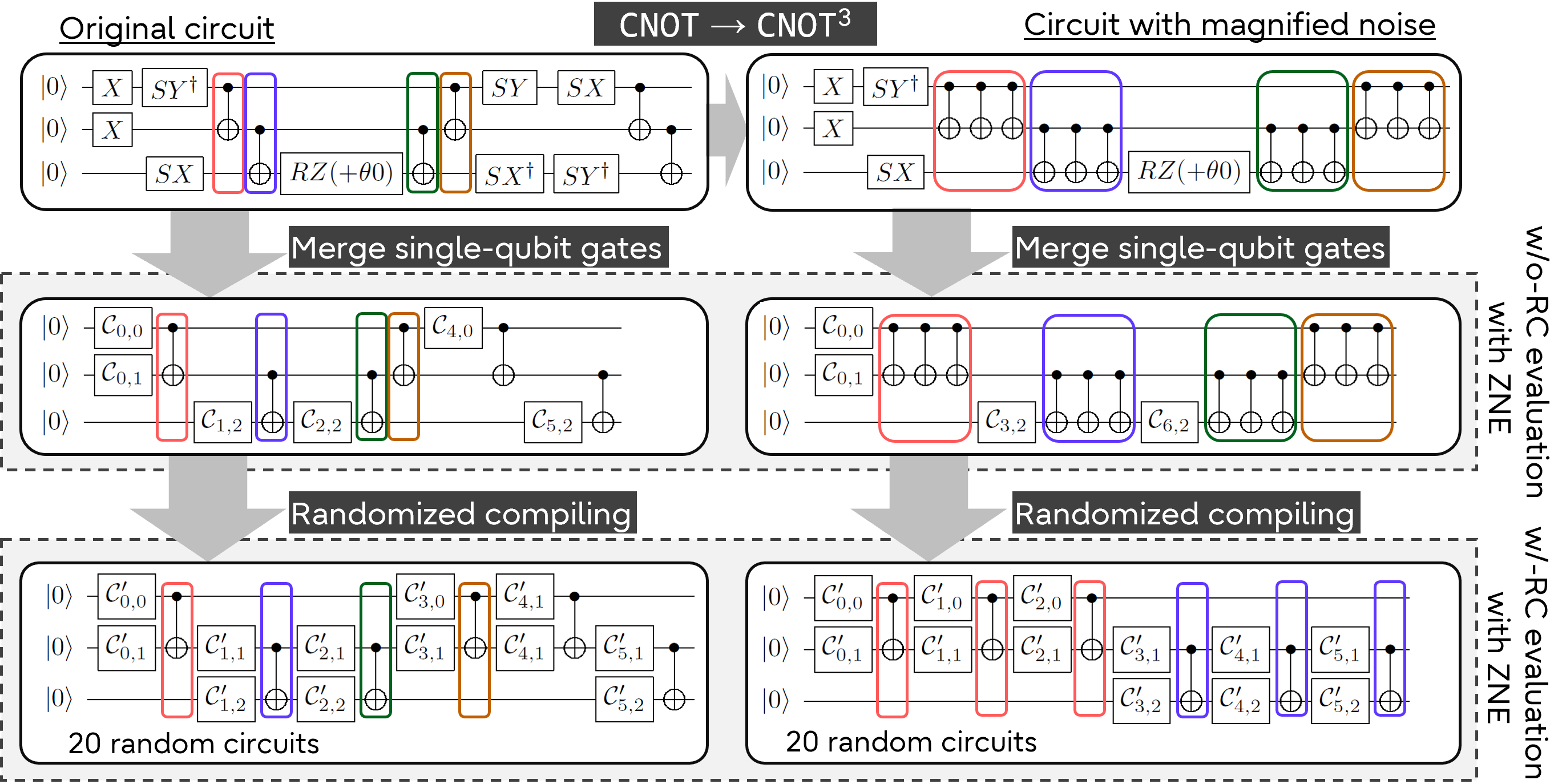}
    \vspace*{-5mm}
    \caption{Schematic illustration of how we perform ZNE with and without RC.  $C_{k,l}$ and $C_{k,l}'$ denote single qubit operations, where k and l denote the cycle index and the qubit index, respectively.}
    \label{fig: zne+rc}
\end{figure*}%

Computing the cost function for a given value of $\vec{\theta}$ can be done using one of two methods: 1) by direct computation of the average energy $\tr{\rho_\theta \mathcal{H}}$, where $\rho_\theta$ is the output state of the circuit, or 2) by computing an empirical average based on a finite number of simulated samples. In the latter case we expand the Hamiltonian as a sum of $n$-qubit Pauli operators 
\begin{align}
\mathcal{H} = \sum_{j = 1}^M a_j P_j
\end{align}
and estimate each term $\tr{\rho_\theta P_j}$ separately by simulating a measurement of the observable $P_j$ using a finite number of shots. Note that the second method represents how VQE has to be implemented in practice, but makes the simulation much slower, since $M$ can be large and shot noise can harm the performance of the classical optimizer. We present results using both methods, showing that the improvement from our error mitigation strategy persists even under the more realistic, but slower, second method. We remark that the measurement overhead is a major bottleneck for VQE algorithms, and in the present work we do not attempt to improve it in our simulations (see \cite{mcclean2016theory, gonthier2022measurements,crawford2021efficient,kurita2022pauli} for various ways to reduce this overhead). 

We perform and compare VQE simulations under various conditions: with no quantum error mitigation, with RC alone, with ZNE alone, and with RC and ZNE combined. The theoretical details of RC and ZNE are described in \cref{apdx: zne,apdx: rc}. The overall VQE scheme with RC and ZNE is depicted in \Cref{fig:overall_scheme}. In this work we utilize the following python modules: True-Q \cite{trueq} for quantum circuit simulation and compilation, SciPy \cite{2020SciPy-NMeth} and BOBYQA \cite{powell2009bobyqa} for classical parameter optimization, and OpenFermion \cite{mcclean2017openfermion} for qubit-mapping of electronic structure Hamiltonians.

For a given value of $\vec{\theta}$, computing the mitigated expected energy under randomized compiling is performed by the following steps (see \Cref{fig: zne+rc}): (i) create noise-magnified version of the ansatz circuit by identity-insertion method (e.g., $\mathtt{CNOT} \mapsto \mathtt{CNOT}^3$). (ii) apply randomized compiling to both circuits. (iii) calculate the expectation values for both circuits, and (iv) apply post-processing to obtain zero-noise estimates of the average energy. 

In Step (ii), randomized compiling is applied in simulation either by replacing the circuit with $N_{\text{rand}}$ randomized compilations (we set $N_{\text{rand}} = 20$ in our experiments, because we find this number to be sufficient for converting coherent errors to Pauli stochastic errors in our simulations \cite{hashim2021randomized}), or by twirling the noise on each noisy clock-cycle, which transforms it to an effective noise process described by a Pauli channel \cite{wallman2016noise}. The latter method has the same effect as performing an exact average of $4^{n k}$ circuits, where $n$ is the number of qubits and $k$ is the number of clock-cycles with two-qubit gates (we abuse terminology and refer to this as the infinite randomization limit), see \Cref{apdx: rc} and Ref.~\cite{wallman2016noise} for more details. In our simulations noise affects only two-qubit gates, which are non-parameterized in our circuit ansatz. Hence we can pre-compute the twirled noise operators outside of the VQE loop, which significantly improves the runtime of the simulation.
We present results using both methods and show that the improvement due to our error mitigation strategy holds in both cases.

\begin{figure*}[t]
    \centering
    \includegraphics[scale = 0.7]{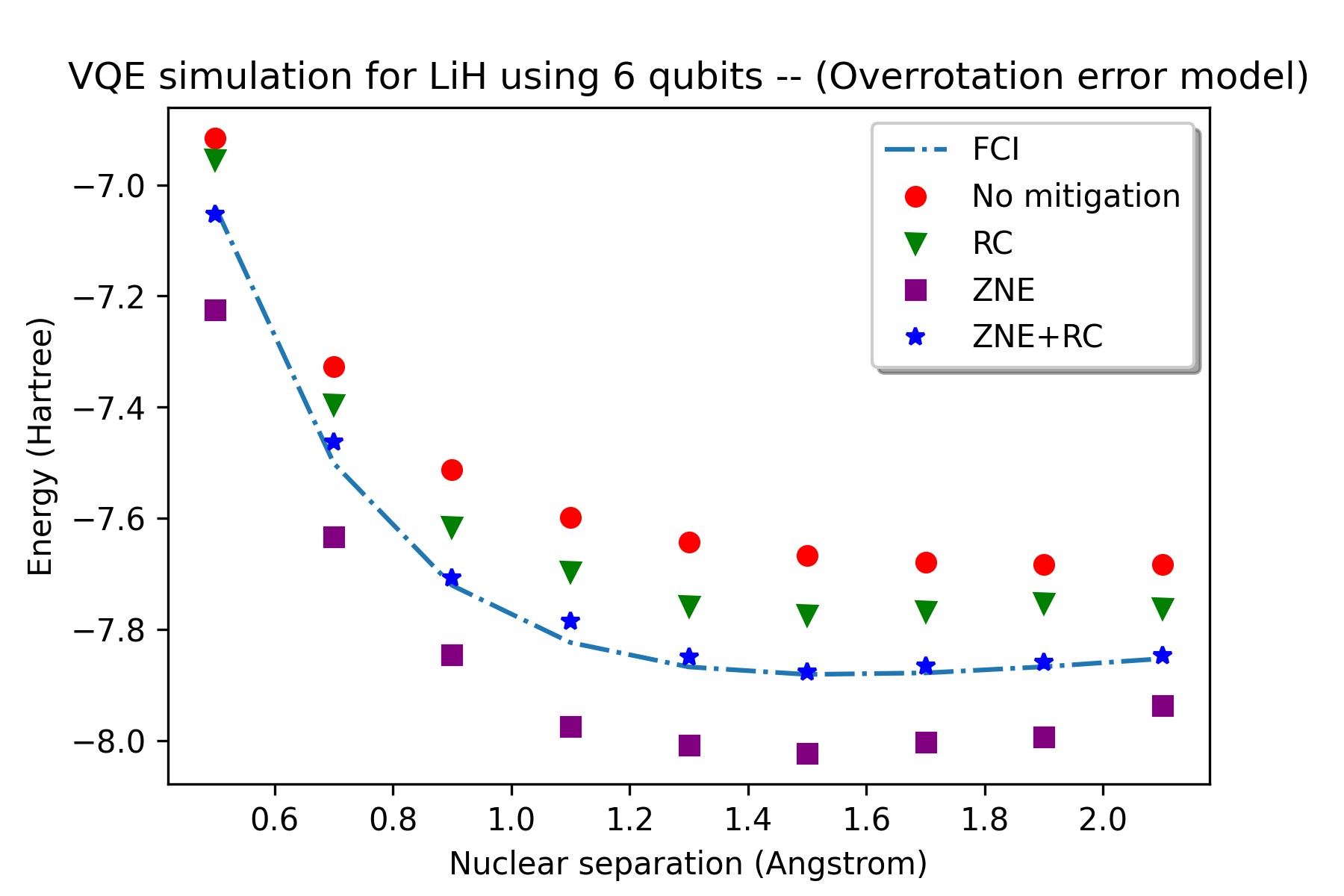}
    \vspace*{-3mm}
    \caption{Theoretical and experimental disassociation curves for the LiH molecule. The theoretical curve is obtained by the full-configuration interaction (FCI) method. The experimental curves are generated by classical simulation of a VQE algorithm on a noisy quantum computer. Here the noise model is the over-rotation model discussed in the main text, and the classical optimizer used is BOBYQA.}
    \label{LiH_curve}
\end{figure*}

\begin{figure*}[t]
    \centering
    \includegraphics[scale = 0.125]{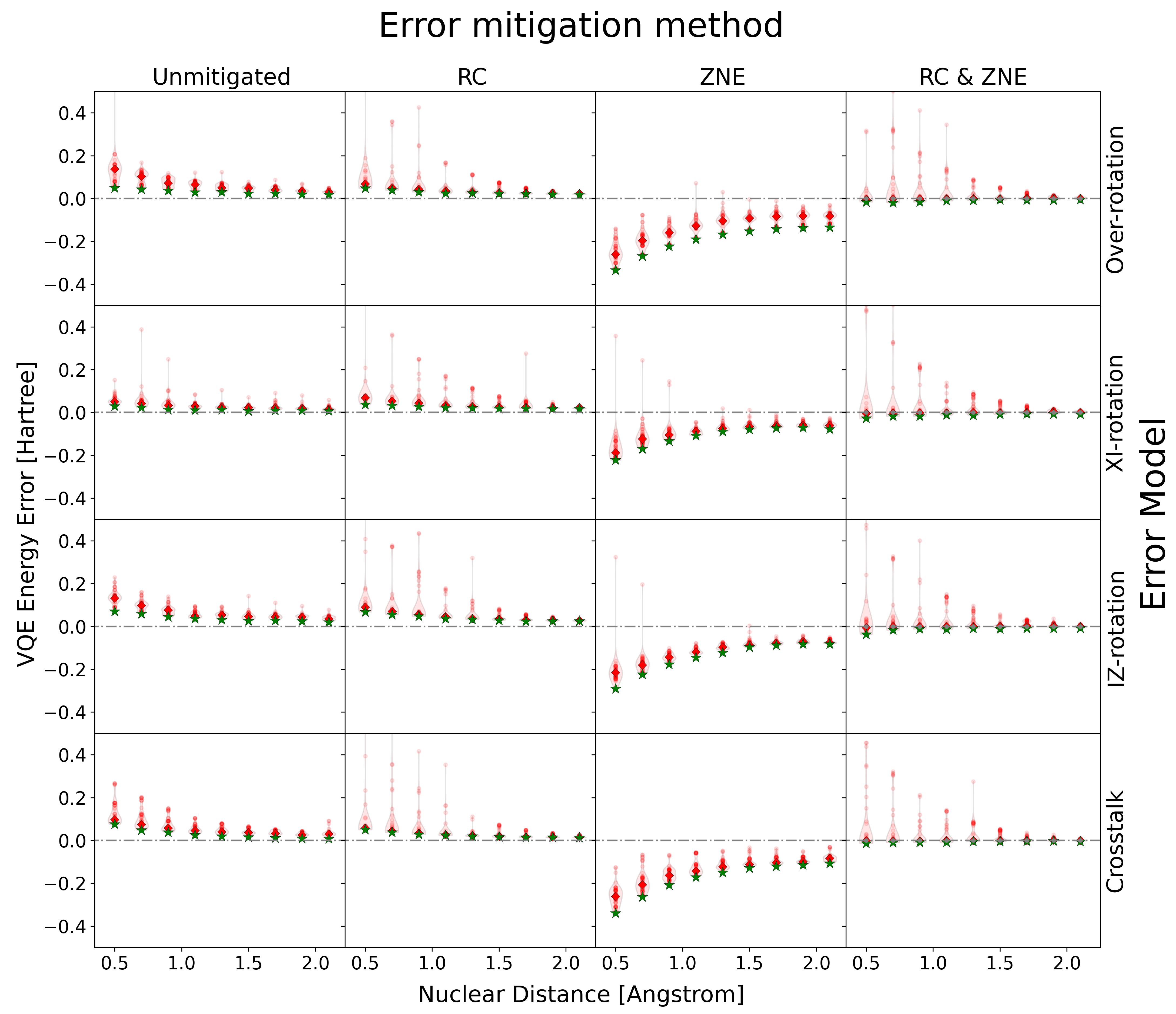}
    \vspace*{-0mm}
    \caption{VQE simulation results for the H$_2$ molecule with the Powell optimizer. The figure shows the energy errors for the energy errors for the four different error models and error mitigation strategies discussed in the text. The energy error is defined as $E_{\text{vqe}} - E_{\text{exact}}$. Red diamonds (green stars) denote median (minimum) energy estimates among 35-60 trials for each molecular configuration. Raw data are shown as open circles, and the their distribution for each molecular configuration is shown in a violin plot.}
    \label{H2}
\end{figure*}

\section{Results and discussion}\label{sec: results}

\begin{figure*}[t]
    \centering
    \includegraphics[scale = 0.125]{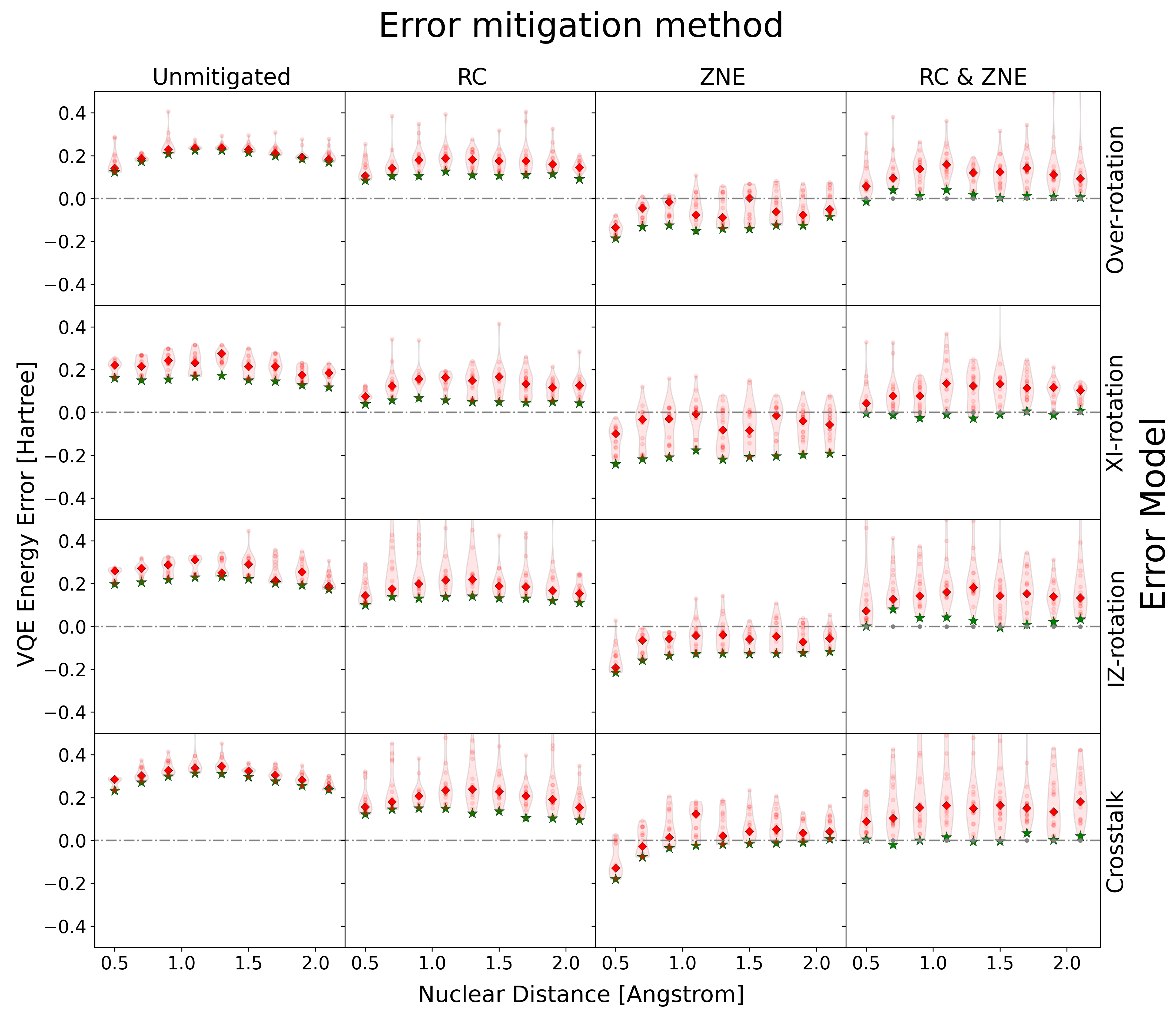}
    \vspace*{-0mm}
    \caption{VQE simulation results for the LiH molecule with the BOBYQA optimizer. The figure shows the energy errors for the four different error models and error mitigation strategies discussed in the text. The energy error is defined as $E_{\text{vqe}} - E_{\text{exact}}$. Red diamonds (green stars) denote median (minimum) energy estimates among 35 trials for each molecular configuration. Raw data are shown as open circles, and the their distribution for each molecular configuration is shown in a violin plot.}
    \label{LiH}
\end{figure*}

Our results indicate significant gains in terms of accuracy by combining RC and ZNE in deep VQE algorithms. 
This is true even though applying each separately is much less effective. \Cref{LiH_curve} shows an example of the effect of error mitigation on simulating a molecular disassociation curve. In all of the numerical experiments we performed a similar behaviour is observed, which seems to be robust to changing various details in the implementation of VQE, such as using different classical optimizers, as well as details of our numerical simulations, such as simulating measurements by sampling vs. directly computing expectation values. 

\Cref{H2} shows the error on the simulated dissociation curve of the H$_2$ molecule via noisy VQE experiments, with and without error mitigation. In this figure (and similar figures below) the energy error is defined as the simulated energy value minus the exact energy value obtained by the full configuration interaction (FCI) method \cite{mcardle2020quantum}, and the violin plots are with respect to 35 – 60 VQE trials with random initial parameters, each trial requiring roughly $100-400$ evaluations of the objective function to reach convergence. 
\Cref{LiH} shows similar results for the LiH molecule. 
In \Cref{LiH} the improvement from combining RC \& ZNE is quite significant when considering the minimum energy amongst a number of VQE trials with random initial parameters. We note that taking the minimum energy amongst a number of trials with random initial points is likely to be the best strategy in any real life application of VQE, since the starting point in parameter space will likely have a significant impact on the convergence of the algorithm.
The results in \Cref{H2,LiH} clearly show that the energy error induced by coherent noise can be significantly mitigated by combining randomized compiling with zero-noise extrapolation. The magnitude of the error reduction depends on the noise model and strength. In our simulation, the energy error can be reduced by 1 – 2 orders of magnitude.

In the simulations in \Cref{H2,LiH}, the cost function is evaluated within the VQE loop by computing the expectation value of the energy exactly for each desired value of the variational parameters (see \Cref{subsec: vqe} for more details.) In real-life implementations of VQE one must estimate the expectation value of the Hamiltonian empirically. We provide results in the appendix (\Cref{H2_sampling}) based on measurement by simulated sampling, and find that a similar improvement from our error mitigation strategy persists. Here we use the simplest measurement strategy where each Pauli term in the Hamiltonian is measured separately. As mentioned previously, we do not attempt to improve the performance of VQE by optimizing the measurement strategy in this work.

\begin{figure}[!t]
    \centering
    \includegraphics[scale = 0.104]{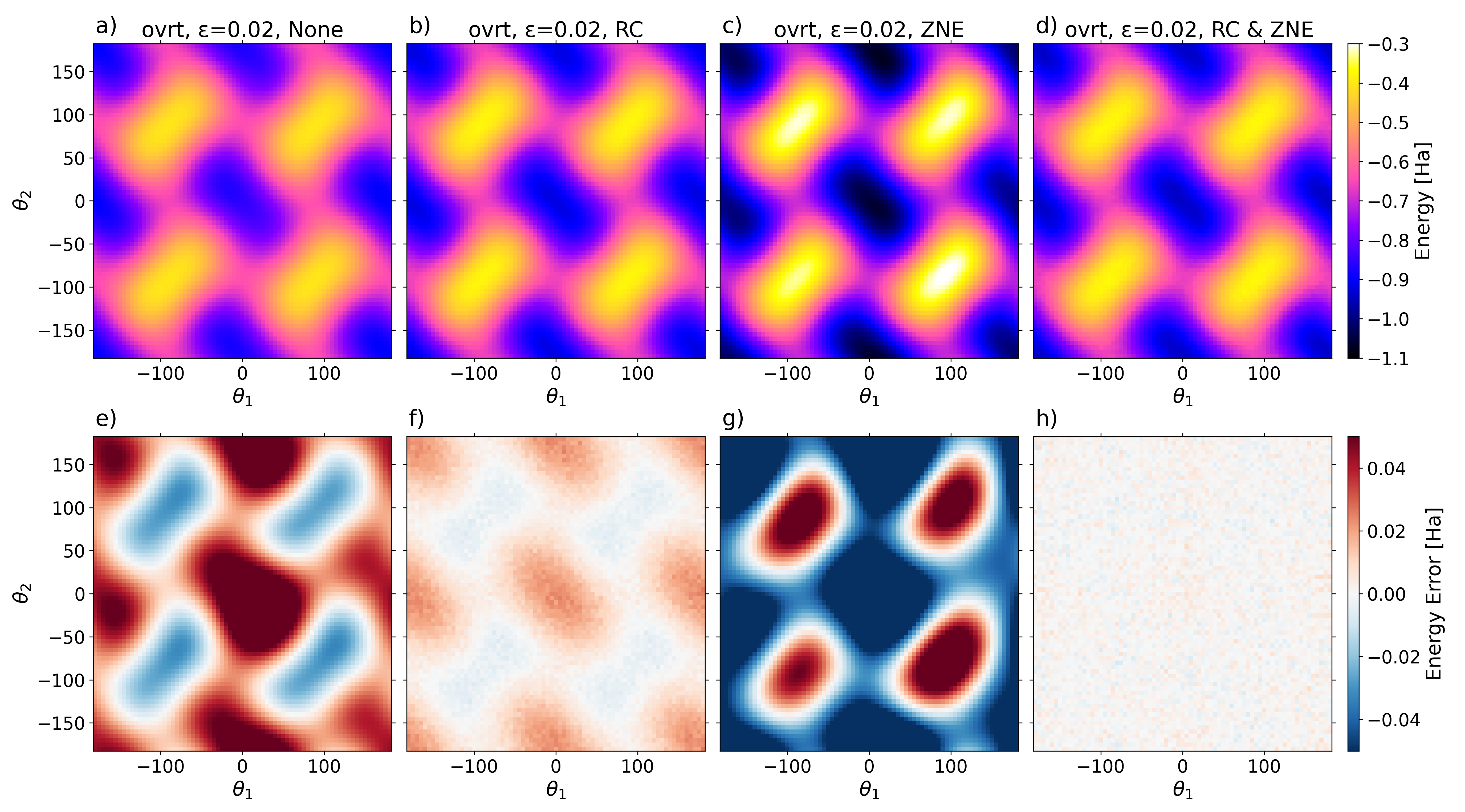}
    \vspace*{-5mm}
    \caption{(a-d) Heatmaps estimated for the energy of H$_2$ under under over-rotation noise ($\epsilon=0.02$), (a) without any quantum error mitigation scheme, (b) with only RC, (c) with only ZNE, and (d) with both RC and ZNE. (e-h) Energy deviation from the ideal heatmap. The first two parameter $\theta_1, \theta_2$ are vairable while $\theta_0$ is fixed at $8.6^\circ$.}
    \label{fig: energy surface}
\end{figure}

We also study the effect of this error mitigation strategy on the energy landscape; that is, the value of the expected energy as a function of the variational parameters. \Cref{fig: energy surface} shows two-dimensional slices of the energy landscape under over-rotation noise and their deviation from the ideal energy. The figure shows that we can drastically reduce the errors induced by coherent noise over the entire energy surface by combining RC and ZNE. Similar data demonstrating the improvement in the error-mitigated energy landscape with other noise models is included in \Cref{apdx figures} (\Cref{fig:heatmap iz,fig:heatmap xi}). 
\Cref{fig: heatmap ovrt id insert} in the same appendix shows that noise magnification by identity insertion not only globally flattens the energy landscape but also modifies its characteristic shape. \Cref{fig: heatmap ovrt id insert w RC} shows the analogous situation under randomized compiling. Contrary to the previous case (\Cref{fig: heatmap ovrt id insert}), with RC noise magnification does not significantly alter the characteristic shape of the landscape. This is a clear manifestation of the noise suppression effect of RC. Adding ZNE to the simulations unleashes the full benefit of our error mitigation strategy, as shown for three different error models in \Cref{fig: energy surface,fig:heatmap xi,fig:heatmap iz}, which demonstrate a remarkable restoration of the energy landscape by combining RC and ZNE.

Our main results show significant improvement in accuracy of the mitigated ground energy, as measured by the minimum optimal value amongst a number of VQE trials with random initial points. In terms of precision, we observe a notable effect where the spread of the optimal values (over random initial points, at any fixed nuclear separation) is significantly increased when using RC and ZNE, most apparently in the violin plots in \Cref{LiH}. This spread indicates the consistency with which the classical optimizer converges to the same solution given random starting points. \Cref{LiH_comparison} shows that this spread strongly depends on the particular classical optimizer, and hence it is in part due to sensitivity of the optimizer to initial conditions. This sensitivity is likely due to a complex interplay between the particular classical optimization routine, the error mitigation method used, and the type and strength of the noise. Indeed, this sensitivity is greatly reduced in our simulation of the infinite-randomization, infinite-shot limit, where each noise source is replaced by its associated Pauli channel under twirling, and the expected energy is computed exactly, see \Cref{LiH_comparison_inf} in the appendix. In this latter figure the variance is almost entirely eliminated by combining ZNE \& RC. This indicates that the performance of the classical optimizers in VQE can be greatly enhanced by carefully choosing hyper-parameters in the error mitigation strategy, such as the number randomizations used in RC. 

To eliminate the influence of the classical optimizer in the above discussion, we fix the variational parameters to some value and consider the energy error in the output state of the resulting circuit.
In \Cref{precision plot} the variance of the estimated energy depends on the error rate and the number of randomizations, with a clear reduction in variance as the number of randomizations increases. We include similar results in \Cref{apdx figures} which demonstrate the same behaviour with other coherent noise models. \Cref{precision 2,precision 3} show histograms of energy deviations from the ideal values for the H$_2$ ansatz for some fixed parameter values. \Cref{precision plot}(d-f), \Cref{precision 2}(d-f), and \Cref{precision 3} (d-f), show that the precision of the energy is proportional to $1/\sqrt{N_{\text{rand}}}$, where $N_{\text{rand}}$ denotes the number of random circuits, confirming a theoretical result \cite{wallman2016noise}.

Finally, to provide some intuition for the significant synergy effect of RC and ZNE, we simulate expectation values of the 14 Pauli terms comprising the H$_2$ Hamiltonian, and we vary the $\mathtt{CNOT}$ infidelity by repeating it an odd number of times, see \Cref{linearized 1,linearized 2,linearized 3} in \Cref{apdx figures}. Without RC the resulting expectation values show strong nonlinear dependence on the number of $\mathtt{CNOT}$ repetitions. This is due to the nature of coherent noise, where errors can accumulate quadratically \cite{wallman2016noise} and in a way that depends on the state. When we apply RC in each simulation the dependence of the resulting expectation values on the number of repetitions becomes almost linear, facilitating extrapolation to the zero-noise limit. Related ideas have recently been suggested \cite{ferracin2022efficiently} and demonstrated in experiment \cite{kim2021scalable,berg2022probabilistic}.

\section{Conclusion and future perspective}\label{sec conclusion}

We propose a quantum error mitigation technique for the VQE algorithm to reduce energy errors induced by coherent noise by using RC and ZNE. We demonstrate that, with this strategy, energy errors in VQE simulations induced by coherent noise are reduced by up to two orders of magnitude. Accordingly, our analysis of the energy landscape shows large improvements when RC and ZNE are combined. This can be explained by the noise-twirling effect of RC, in which coherent noise is converted to stochastic noise, resulting in a more linearized dependence of expectation values with gate infidelities, enabling ZNE to successfully extrapolate to the zero-noise limit. This proposed error mitigation technique can be used for other quantum algorithms which involve calculating expectation values of observables.

In future research, more general quantum noise (including relaxation and measurement noise) will be applied to investigate the efficacy of this error mitigation strategy in a more practical setting. Also, the identity insertion method which we employed here is not suitable for experiments with current NISQ devices because the insertion increases the depth of the corresponding circuit by a factor of three or more. This problem may be solved by randomly choosing some entangling gates to magnify their noises instead of magnifying all the entangling gates \cite{he2020zero}. 

\bibliography{references.bib}
\bibliographystyle{quantum}

\onecolumn
\appendix

\section{Background theory}

\subsection{Variational quantum eigensolver and unitary coupled cluster ansatz}\label{appdx: vqe}

The VQE algorithm is a quantum-classical hybrid algorithm. The quantum part is used to evaluate the cost function by measuring certain expectation values in candidate parameterized quantum states, and the classical part performs the optimization, updating the parameters and calling the quantum part as a subroutine. 
The goal of VQE experiments is to approximate the ground state energy of a Hamiltonian $\mathcal{H}$. In this work $\mathcal{H}$ represents the electronic structure Hamiltonian of some molecule.
More specifically, the nuclei in the molecule are represented by stationary, classical point charges (the Born-Oppenheimer approximation). Under this approximation we obtain a second quantized Hamiltonian of the form

\begin{align}
\mathcal{H} = \sum_{p, q} h_{pq} a_p a_q^\dagger + \sum_{p, q, r, s} h_{pqrs} a_p a_q a_r^\dagger a_s^\dagger,
\end{align}
where $a_i$ and $a_i^\dagger$ denote particle-annihilation and -creation operators on the $i$-th spin orbital, the summation indices $\{p, q, r, s\}$ run over a set of electronic spin-orbitals, and $h_{pq}$ and $h_{pqrs}$ are coefficients that can be obtained by standard chemistry calculations. The goal of the VQE experiments in this paper is to study the dependence of the ground state energy on the positions of the nuclei, and reproduce the so-called potential energy surface of the molecule (or a one-dimensional slice of it, commonly known as a disassociation curve).

In this paper, we employ the STO-3g basis set and the unitary coupled-cluster ansatz with single- and double-excitations (UCC-SD) \cite{mcclean2016theory}. UCC-SD is a structured ansatz which takes advantage of the details of the Hamiltonian and the physics describing the system, associating to each variational parameter a particle excitation from a reference state, such as the Hartree-Fock (HF) state, to capture additional electronic correlations not present in HF. In contrast to unstructured ansatz, such as the hardware-efficient ansatz \cite{kandala2017hardware}, UCC-SD requires much fewer parameters, but requires additional circuit depth. This is preferable for our purpose since it avoids energy errors caused by optimization failure due to the number of parameters being too large -- classical optimization with fewer parameters is likely to result in more exact outcomes. For this ansatz, we choose an initial quantum state which encodes the Hartree-Fock solution $\ket{\psi_{\mathrm{HF}}}$, and apply a quantum circuit which approximates the state 
\begin{align}
\ket{\psi(\theta)} = e^{T(\theta)-T(\theta)^\dagger} \ket{\psi_{\mathrm{HF}}}
\end{align}
by Trotterization,
where
\begin{align}\label{cluster operator}
T(\theta) = \sum_{i \in S_{virt}}\sum_{k \in S_{occ}} \theta_{i, k} a_i^\dagger a_k + \sum_{i, j \in S_{virt}} \sum_{k, \ell \in S_{occ}} \theta_{i, j, k, \ell} a_i^\dagger a_j^\dagger a_k a_\ell,
\end{align}
where $a_i^\dagger$ denotes a particle-creation operator on the $i$-th spin-orbital, and $a_k$ denotes a particle-annihilation operator on the $k$-th spin-orbital. Here the sets $S_{virt}$ and $S_{occ}$ denote the subsets of occupied and unoccupied spin-orbitals in some reference Slater determinant, which we take to be the Hartree-Fock state. The first term in \cref{cluster operator} denotes single-particle excitation operators with amplitude $\theta_{i, k}$, and the second term denotes two-particle excitation operators with amplitude $\theta_{i,j,k,l}$. We regard $\theta_{i,k}$ and $\theta_{i,j,k,l}$ as the parameters for optimization. For example, when the target molecule is H$_2$, there are two parameters for single-particle excitation (which we denote $\theta_0,\theta_1$), and one parameter for two-particle excitation (which we denote $\theta_2$). Under the Jordan-Wigner transformation, $a_i^\dagger$ and $a_k$ is transformed to $(1/2) Z_0 \cdots Z_{i-1} (X_i-iY_i)$ and $(1/2) Z_0 \cdots Z_{k-1} (X_k+iY_k)$, respectively. We employ Trotter decomposition to implement this Hamiltonian evolution $e^{T-T^\dagger}$ as a quantum operation circuit.

\subsection{Zero-noise extrapolation}\label{apdx: zne}
Here we provide an overview of ZNE \cite{li2017efficient,temme2017error,he2020zero}. Suppose the gate error rate is $\epsilon$, and the expectation value calculated with the gate error $\epsilon$ is $E(\epsilon)$. We can magnify such gate error to $r\epsilon$ (where $r>1$) by lengthening the duration of target operation or inserting operations equivalent to identities $(U^\dagger U)$ prior to the target operation $(U)$. We suppose $E(r\epsilon)$ can be polynomial of $r\epsilon$:
\begin{align}
E(r\epsilon)=E(0)+ \lambda_1 r\epsilon+ \lambda_2 (r\epsilon)^2+O((r\epsilon)^3 )
\end{align}
When we observe $E$ with two noise magnification coefficients $r_1,r_2$  (typically $r_1=1$), we can do first-order correction of $E$: 
\begin{align}
E^* (0)=r_2/(r_2-r_1 ) E(r_1 \epsilon)-r_1/(r_2-r_1 ) E(r_2 \epsilon)+O(\epsilon^2)
\end{align}
Such extrapolation method is valid when $\epsilon$ is sufficiently small. 
The accuracy of $E^* (0)$ naturally depends on that of $E(r_1 \epsilon)$ and $E(r_2 \epsilon)$, which technically depend on the number of shots. Also, large absolute values of $r_2/(r_2-r_1 )$ and $r_1/(r_2-r_1 )$ increase the accuracy of $E^*(0)$.

\subsection{Randomized compiling}\label{apdx: rc}
As briefly described in \Cref{sec intro}, RC is a quantum error mitigation technique that utilizes Pauli twirling. To execute randomized compiling properly, we describe the overall quantum operation $U$ as
\begin{align}
\mathcal{U} \rho = \mathcal{C}_n \mathcal{D}_n \mathcal{C}_{n-1} \mathcal{D}_{n-1} \mathcal{C}_{n-2} \cdots \mathcal{C}_1 \mathcal{D}_1 \mathcal{C}_0 \rho
\end{align}
by employing Pauli transfer matrix representation, where $\rho$ is a $4^N \times 1$ vector which denotes the initial quantum state ($N$: the number of qubits) and $\mathcal{D}_1, \cdots ,\mathcal{D}_n$ and $\mathcal{C}_0,\mathcal{C}_1,..., \mathcal{C}_n$ are $4^N \times 4^N$ matrices which denote the ideal quantum gate operations. We assume that $\mathcal{D}_1, \cdots, \mathcal{D}_n$ causes more noise than $\mathcal{C}_0, \mathcal{C}_1,\cdots, \mathcal{C}_n$ when implemented. As a typical approach to satisfy this assumption, we categorize $\mathcal{D}_1, \cdots, \mathcal{D}_n \in \boldsymbol{\mathcal{D}}$ as the operations with entangling gates, and $\mathcal{C}_0,\mathcal{C}_1, \cdots, \mathcal{C}_n \in \boldsymbol{\mathcal{C}}$ as the operations with single-qubit gates \cite{wallman2016noise}. To execute RC, we make several equivalent circuits with Pauli twirling, where we change $\mathcal{C}_k$ to $\mathcal{C}'_k \in \boldsymbol{\mathcal{C}}$ while $\mathcal{D}_k$ is unchanged. To do this, for each $k \in \{1, \cdots, n\}$, we insert random Pauli operation $\mathcal{T}_k$ prior to $\mathcal{D}_k$ and compensation operation $\mathcal{T}'_k$ posterior to $\mathcal{D}_k$ so that $\mathcal{D}_k = \mathcal{T}_k \mathcal{D}_k \mathcal{T}'_k$. When we make a new quantum circuit equivalent to the original one with Pauli twirling, we can describe it as
\begin{align}\label{Noiseless Pauli-twirl}
\mathcal{U} \rho 
&= \mathcal{C}_n \mathcal{T}_n \mathcal{D}_n \mathcal{T}'_n \mathcal{C}_{n-1} \mathcal{T}_{n-1} \mathcal{T}_{n-1} \mathcal{D}_{n-1} \mathcal{T}'_{n-1} \mathcal{C}_{n-2} \cdots \mathcal{C}_{1} \mathcal{T}_{1} \mathcal{D}_{1} \mathcal{T}'_{1} \mathcal{C}_{0} \rho \nonumber\\
&= \mathcal{C}'_n \mathcal{D}_n \mathcal{C}'_{n-1} \mathcal{D}_{n-1} \mathcal{C}'_{n-2} \cdots \mathcal{C}'_1 \mathcal{D}_1 \mathcal{C}'_{0} \rho,
\end{align}
where we denote $\mathcal{C}'_k = \mathcal{T}'_{k+1} \mathcal{C}_k \mathcal{T}_{k} (k=1, \cdots, n-1)$, $\mathcal{C}'_0 = \mathcal{T}'_1 \mathcal{C}_0$, and $\mathcal{C}'_n = \mathcal{C}_n \mathcal{T}_n$. 
The requirement that $\mathcal{C}'_{k} \in \boldsymbol{\mathcal{C}}$ for any $k$ and $\mathcal{T}_k$ is satisfied if $\mathcal{D}_k$ is an operation composed of Clifford entangling gates. 

When we describe noisy version of the quantum circuit, it can be written as
\begin{align}
\tilde{\mathcal{U}} \rho 
&= \tilde{\mathcal{C}}_n \tilde{\mathcal{D}}_n \tilde{\mathcal{C}}_{n-1} \tilde{\mathcal{D}}_{n-1} \tilde{\mathcal{C}}_{n-2} \cdots \tilde{\mathcal{C}}_1 \tilde{\mathcal{D}}_1 \tilde{\mathcal{C}}_0 \rho \nonumber\\
&= \mathcal{C}_n \mathcal{N}_{\mathcal{C}_n} \mathcal{N}_{\mathcal{D}_n} \mathcal{D}_n \mathcal{C}_{n-1}\mathcal{N}_{\mathcal{C}_{n-1}} \cdots \mathcal{C}_1 \mathcal{N}_{\mathcal{C}_1} \mathcal{N}_{\mathcal{D}_1} \mathcal{D}_1 \mathcal{C}_0 \mathcal{N}_{\mathcal{C}_0} \rho
\end{align}
where $\mathcal{N}_{\mathcal{C}_k} (k = 0, \cdots, n)$ and $\mathcal{N}_{\mathcal{D}_k}$ $(k= 1,\cdots, n)$ denote the noise operators corresponding to $\mathcal{C}_k$ and $\mathcal{D}_k$, and $\tilde{\mathcal{C}}_k = \mathcal{C}_k \mathcal{N}_{\mathcal{C}_k}$ and $\tilde{\mathcal{D}}_k = \mathcal{N}_{\mathcal{D}_k} \mathcal{D}_k$.
Here, we can set $\mathcal{N}_{C_k} = \mathcal{N}_{\mathcal{C}}$ for any $\mathcal{C}_k \in \boldsymbol{\mathcal{C}}$ with the assumption that the noise amplitude fluctuations among $\mathcal{C}_k \in \boldsymbol{\mathcal{C}}$ are negligible compared to the noise amplitudes of $\mathcal{N}_{\mathcal{D}_1}, \cdots, \mathcal{N}_{\mathcal{D}_n}$.

We can similarly describe the noisy version of Pauli-twirled circuit (\cref{Noiseless Pauli-twirl}) as
\begin{align}
\tilde{\mathcal{U}}^{\mathrm{PT}} \rho 
&= \mathcal{C}'_n \mathcal{N}_{\mathcal{C}} \mathcal{N}_{\mathcal{D}_n} \mathcal{D}_n \mathcal{C}'_{n-1} \mathcal{N}_{\mathcal{C}} \cdots \mathcal{C}'_1 \mathcal{N}_{\mathcal{C}} \mathcal{N}_{\mathcal{D}_1} \mathcal{D}_1 \mathcal{C}'_{0} \mathcal{N}_{\mathcal{C}} \rho \nonumber\\
&= \mathcal{C}_n \mathcal{T}_n \mathcal{N}_{\mathcal{C}} \mathcal{N}_{\mathcal{D}_n} \mathcal{D}_n \mathcal{T}'_{n} \mathcal{C}_{n-1} \mathcal{T}_{n-1} \mathcal{N}_{\mathcal{C}} \cdots \mathcal{C}_1 \mathcal{T}_1 \mathcal{N}_{\mathcal{C}} \mathcal{N}_{\mathcal{D}_1} \mathcal{D}_1 \mathcal{T}'_1 \mathcal{C}_0 \mathcal{N}_{\mathcal{C}} \rho. 
\end{align}
By approximating the resultant quantum states for all possible twirling operations $\mathcal{T}_1, \dots ,\mathcal{T}_n \in \boldsymbol{\mathcal{T}}$, we obtain
\begin{align}\label{Noise approximation}
\frac{1}{|\boldsymbol{\mathcal{T}^n}|} \sum_{\mathcal{T}_1, \dots, \mathcal{T}_n \in \boldsymbol{\mathcal{T}}} \tilde{\mathcal{U}}_{\mathcal{T}}^{\mathrm{PT}}(\rho) = \frac{1}{|\boldsymbol{\mathcal{T}^n}|} \left[ \prod_{k=n, \dots, 1} \mathcal{C}_k \left(  \sum_{\mathcal{T}_k \in \boldsymbol{\mathcal{T}}} \mathcal{T}_k \mathcal{N}_{\mathcal{C}} \mathcal{N}_{\mathcal{D}_k} \mathcal{T}_k\right) \mathcal{D}_k \right] \mathcal{C}_0 \mathcal{N}_{\mathcal{C}} \rho.
\end{align}
\color{black}This means that every noisy operator $\mathcal{N}_{\mathcal{C}} \mathcal{N}_{\mathcal{D}_k}$ is twirled by $\mathcal{T}_k$ so that the noise corresponding to $\mathcal{D}_k$ is converted to stochastic Pauli noise. Although \cref{Noise approximation} means applying RC requires $|\mathcal{T}|^n$ compiled circuits, we can obtain such effect with fewer compiled circuits \cite{hashim2021randomized,ville2021leveraging,kim2021scalable} by sacrificing the precision of the resultant expectation values \cite{ville2021leveraging}.

In this study, we compile the ansatz circuits so that all $\mathcal{D}_k$ are composed of $\mathtt{CNOT}$ gates for the following reasons: (i) $\mathtt{CNOT}$ is one of the Clifford gates, so if $\mathcal{T}_k$ is an operation with Pauli gates, $\mathcal{T}'_k$ is also one with Pauli gates. (ii) $\mathtt{CNOT}$ is Hermitian, so, for applying ZNE, we can magnify the gate noises by inserting two $\mathtt{CNOT}$ as an identity. We can also choose Clifford but non-Hermitian gates (such as $\mathtt{iSWAP}$) as $\mathcal{D}_k$. In this case, we need more than two gate insertions to magnify gate noises (for example, four $\mathtt{iSWAP}$s are needed to insert as an identity).

\clearpage

\section{Simulation results with more realistic noise models}\label{apdx realnoise}

In the main text, the simulation data with only coherent noises which are affected on entangling gates are shown. However, the realistic noise model can be including (i) single-qubit gate noise and (ii) incoherent noise. In this section, we examine how our error-mitigation protocol works in the condition above.

\subsection{Simulations with noisy single-qubit gates}\label{apdx singleQ}
\begin{figure}[h]
    \centering
    \includegraphics[scale = 0.125]{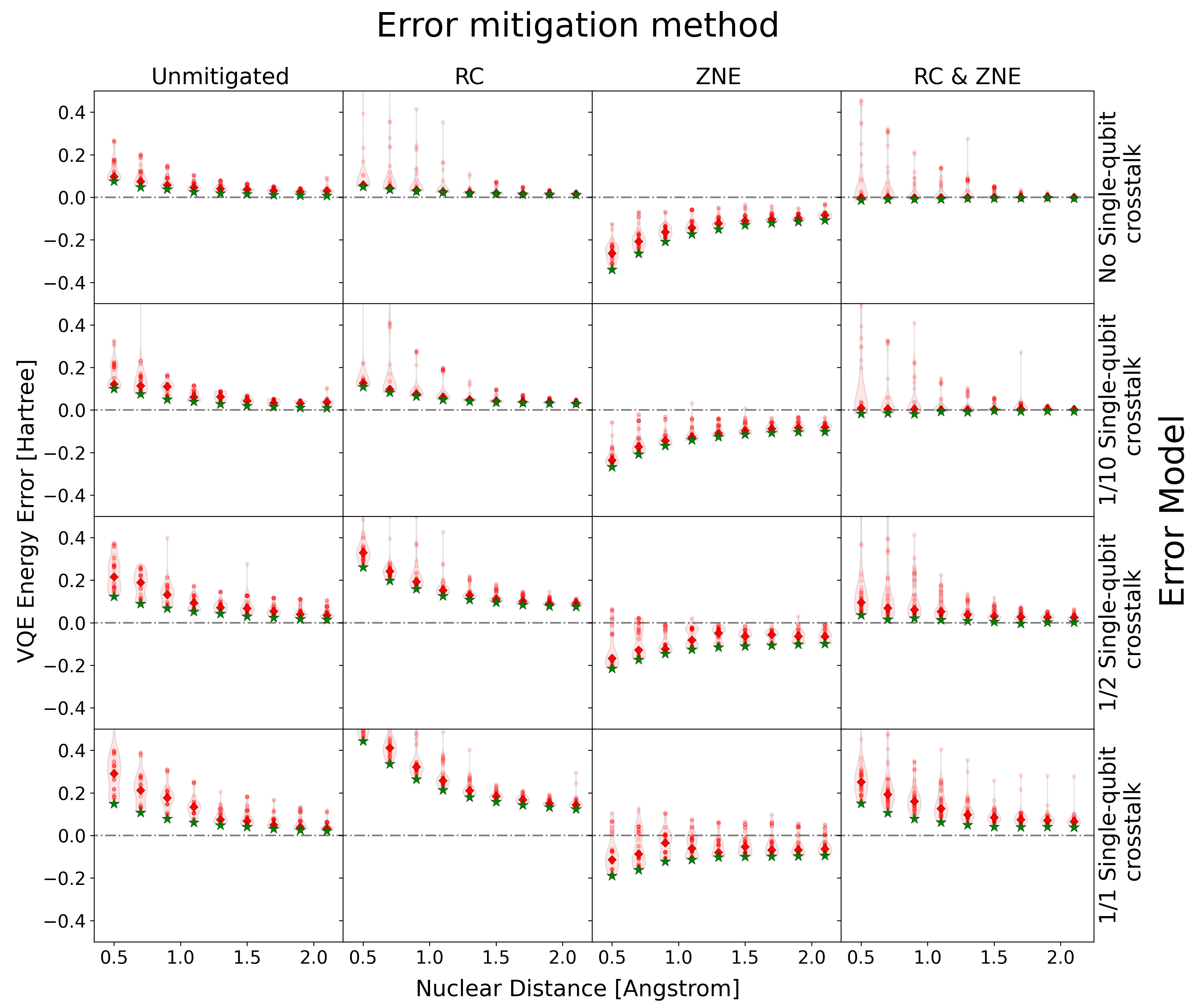}
    \vspace*{-0mm}
    \caption{VQE simulation results for the H$_2$ molecule with the Powell optimizer. The figure shows the energy errors with the coherent crosstalk affected on both single-qubit gates and two-qubit gates. We assume four cases: (i) No single-qubit gate noise, (ii-iv) existance of single-qubit gate noises whose error rate is 1/10, 1/2, and 1/1 of that of entangling gate noises, respectively. We fixed the error rate of entangling gates to $10^{-3}$. We apply four error mitigation strategies discussed in the text. The energy error is defined as $E_{\text{vqe}} - E_{\text{exact}}$. Red diamonds (green stars) denote median (minimum) energy estimates among 60 trials for each molecular configuration. Raw data are shown as open circles, and the their distribution for each molecular configuration is shown in a violin plot.}
    \label{H2_singleQ}
\end{figure}

While error rates on single-qubit gates are typically much smaller than that of entangling gates in most types of hardware, their overall contribution will not be negligible when the number of single-qubit gates becomes large. Furthermore, since randomized compiling will introduce additional single-qubit gates, it is important to investigate how our error mitigation strategy performs when the single-qubit gates are themselves noisy. In this section, we include simulation results that show the resilience of the proposed error mitigation method with noisy single-qubit gates subject to crosstalk. As shown in \Cref{H2_singleQ}, the error mitigation effect with our protocol is still significant when the error rate of single-qubit gates is at most 1/2 that of entangling gates. As the error rate of single-qubit gates gets larger, we find that the resulting energy error cannot be reduced with RC and ZNE.

\subsection{Simulations with incoherent noise}\label{apdx incoherent}
\begin{figure}[h]
    \centering
    \includegraphics[scale = 0.125]{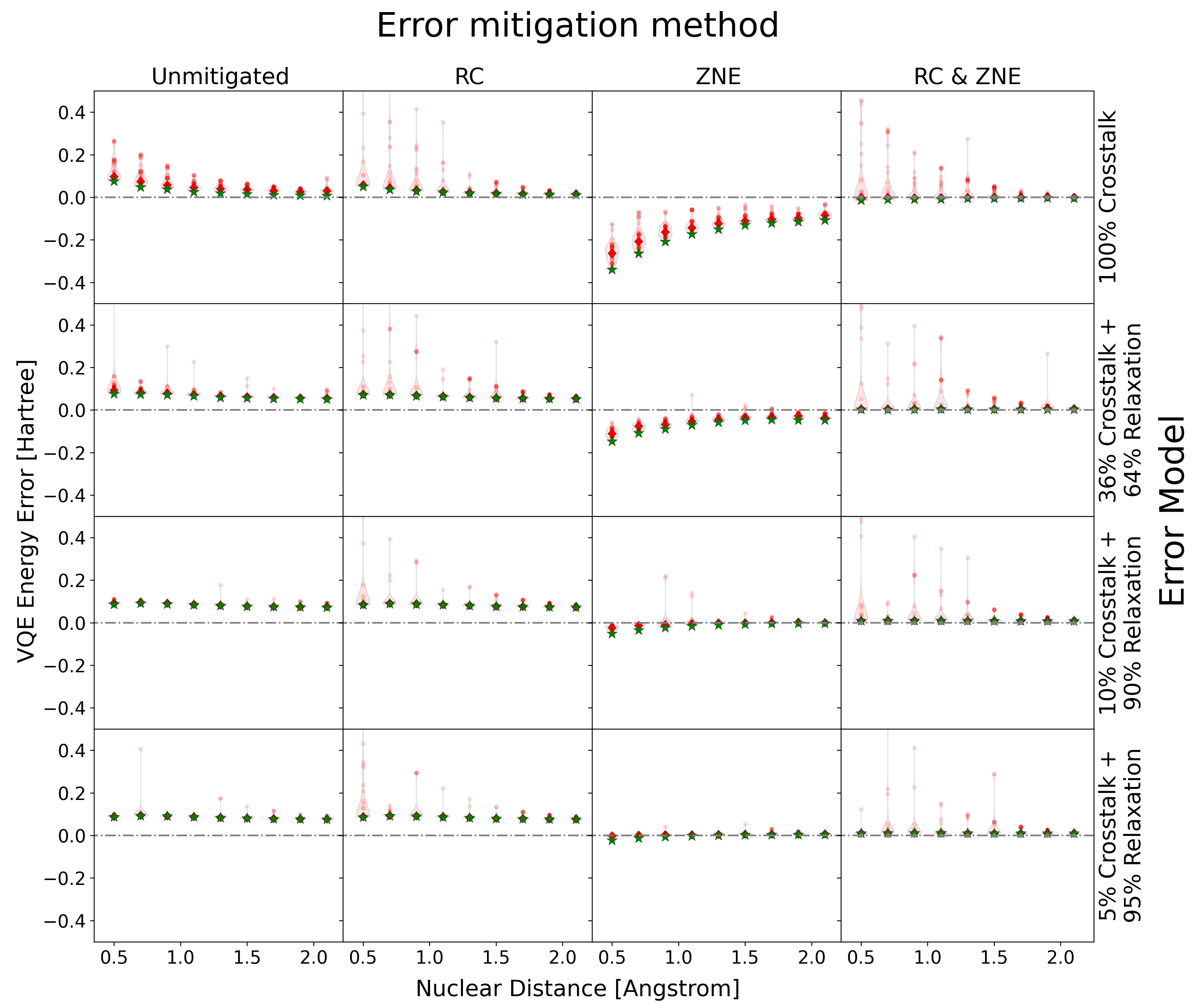}
    \vspace*{-0mm}
    \caption{VQE simulation results for the H$_2$ molecule with the Powell optimizer. The figure shows the energy errors with the relaxation noise as well as the coherent crosstalk, both affected on entangling gates. We assume various ratios of contributions of these error models: (i) 100\%-0\%, (ii) 36\%-64\%, (iii) 10\%-90\%, (iv) 5\%-95\%. We fixed the overall error rate of entangling gates to $10^{-3}$. We apply four error mitigation strategies discussed in the text. The energy error is defined as $E_{\text{vqe}} - E_{\text{exact}}$. Red diamonds (green stars) denote median (minimum) energy estimates among 60 trials for each molecular configuration. Raw data are shown as open circles, and the their distribution for each molecular configuration is shown in a violin plot.}
    \label{H2_incoherent}
\end{figure}

In the main text, we show simulation results with coherent noise only. In this section we include simulations with incoherent noise, in the form of a generalized qubit relaxation process described by a Choi matrix
\begin{align}
C=
\begin{bmatrix}
1 - p( 1 - e ^ { -t / T_1 } ) & 0                          & 0                  & e ^ {-t / T_2} \\
0                             & ( 1 - p ) e ^ { -t / T_1 } & 0                  & 0              \\
0                             & 0                          & p e ^ { -t / T_1 } & 0              \\
e ^ {-t / T_2}                & 0                          & 0            & 1 - ( 1 - p ) e ^ { -t / T_1 }
\end{bmatrix}
\end{align}
with $T_1 = 10$, $T_2 \in [ 1.73 , 5.00] $ and $p=0$. In these simulations we choose gate times $t = 6 \times 10 ^ {-4}$. Since RC works only on coherent noise, we expect less error-reduction effect by employing RC alone when the contribution of coherent noises gets smaller. To confirm this, we show the simulation results under the existence of both coherent (crosstalk) and incoherent (generalized relaxation) noise. As shown in \Cref{H2_incoherent}, when we employ only ZNE, the resulting errors are reduced as the contribution of coherent noise decreases. However, the results show that the error mitigation effect with both RC and ZNE exists even when the contribution of coherent noise is only 10\% of the overall error rate.

\clearpage

\section{Figures and simulation results}\label{apdx figures}

\begin{figure}[h]
    \centering
    \includegraphics[scale = 0.125]{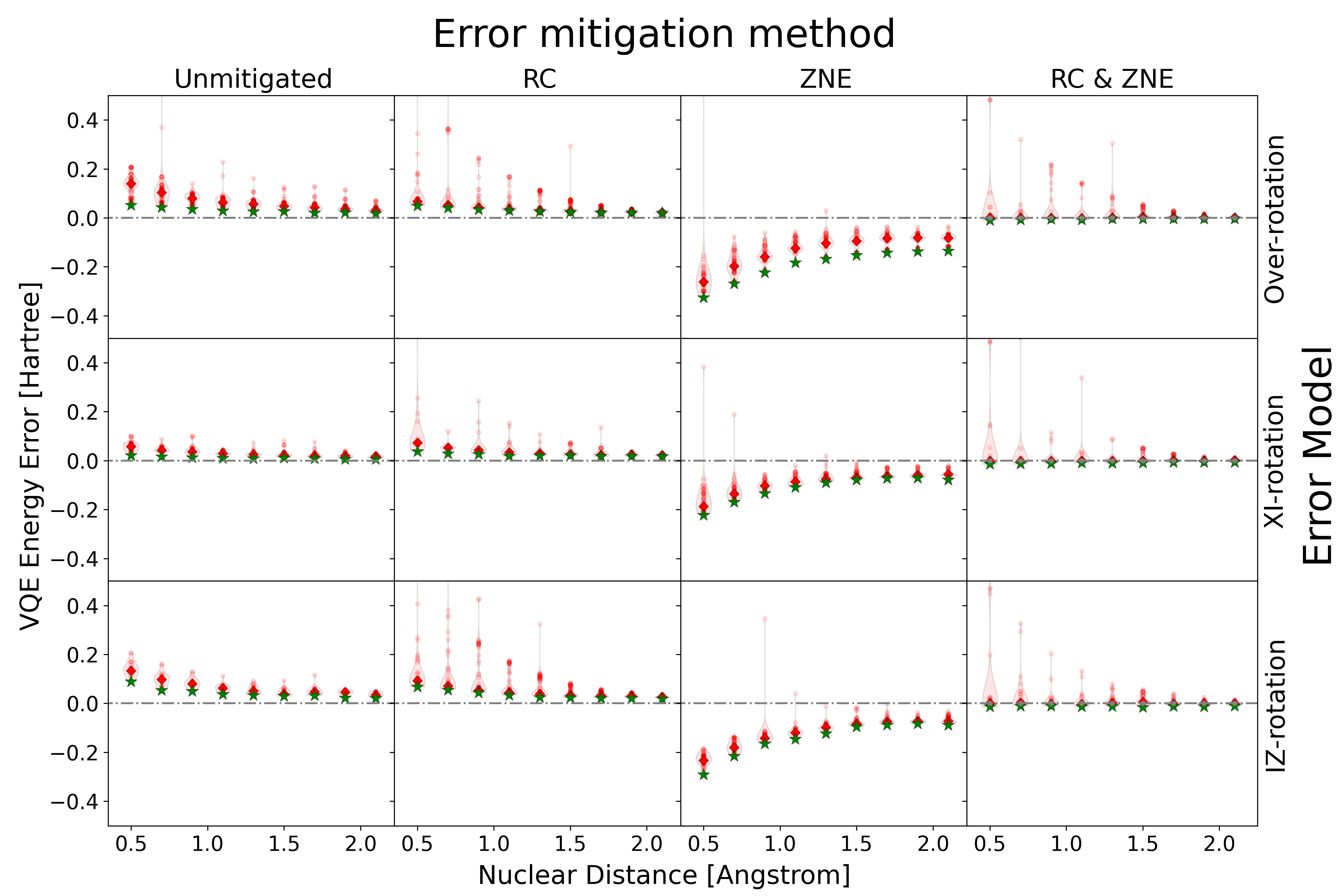}
    \vspace*{-0mm}
    \caption{VQE simulation results for the H$_2$ molecule with the Powell optimizer. The figure shows the energy errors for the four different error models and error mitigation strategies discussed in the text. The energy error is defined as $E_{\text{vqe}} - E_{\text{exact}}$. Red diamonds (green stars) denote median (minimum) energy estimates among 35-60 trials for each molecular configuration. Raw data are shown as open circles, and the their distribution for each molecular configuration is shown in a violin plot. Unlike \Cref{H2}, where the cost function in the VQE simulation is computed using linear algebra, in the above simulation the cost function in VQE is computed by simulated sampling for measuring each of the 14 Pauli term in the H$_2$ Hamiltonian.}
    \label{H2_sampling}
\end{figure}

\begin{figure}[!htb]
    \centering
    \includegraphics[scale = 0.104]{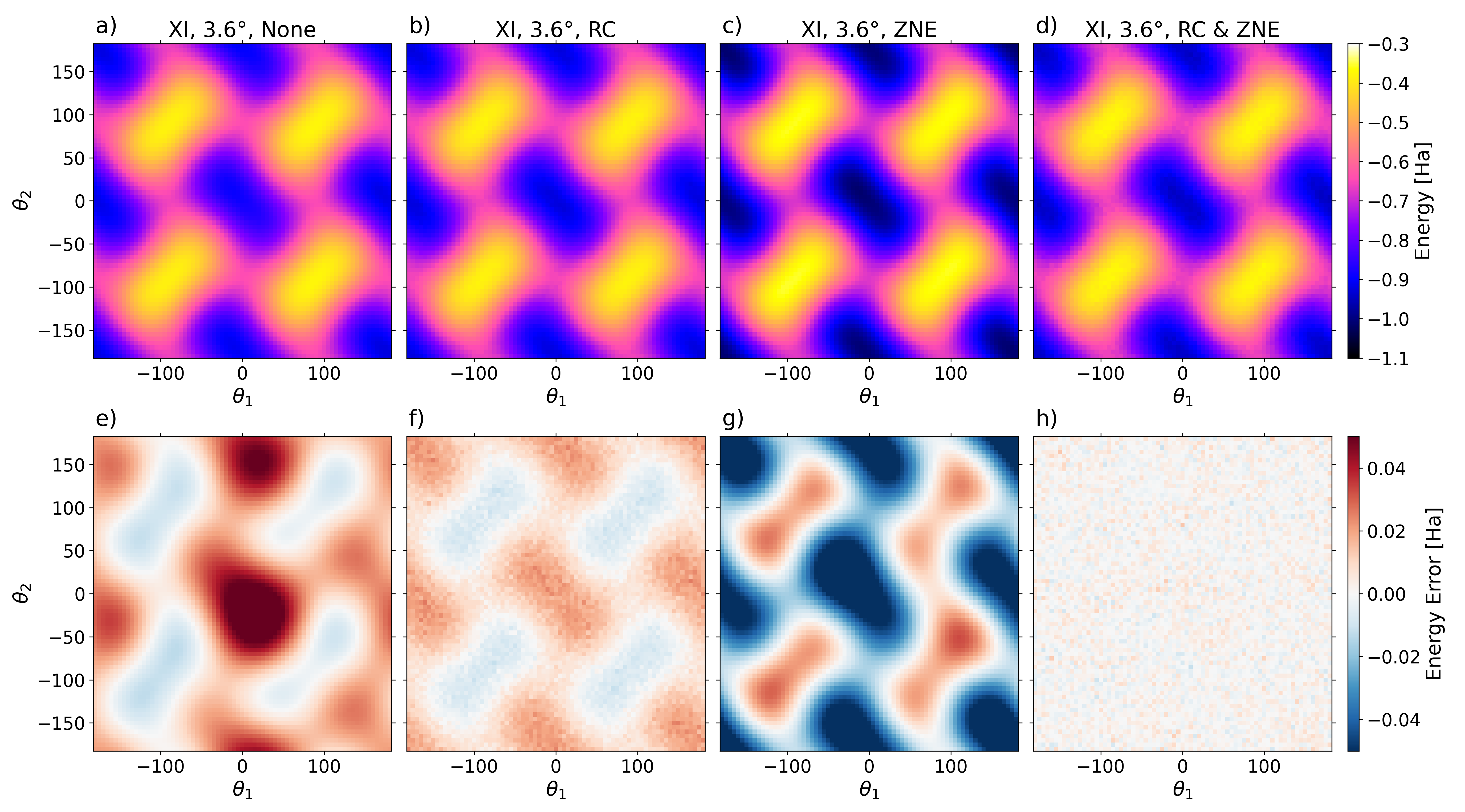}
    \vspace*{-5mm}
    \caption{(a-d) Heatmaps estimated via noisy quantum operation under XI-rotation noise $(\phi=3.6^\circ)$, (a) without any quantum error mitigation scheme, (b) with only RC, (c) with only ZNE, and (d) with both RC and ZNE. (e-h) Deviation heatmaps from the ideal heatmap in \Cref{fig: heatmap ovrt id insert}(a) for (a-d). The first two parameter $\theta_1, \theta_2$ are vairable while $\theta_0$ is fixed at $8.6^\circ$.}
    \label{fig:heatmap xi}
\end{figure}%

\begin{figure}[!htb]
    \centering
    \includegraphics[scale = 0.104]{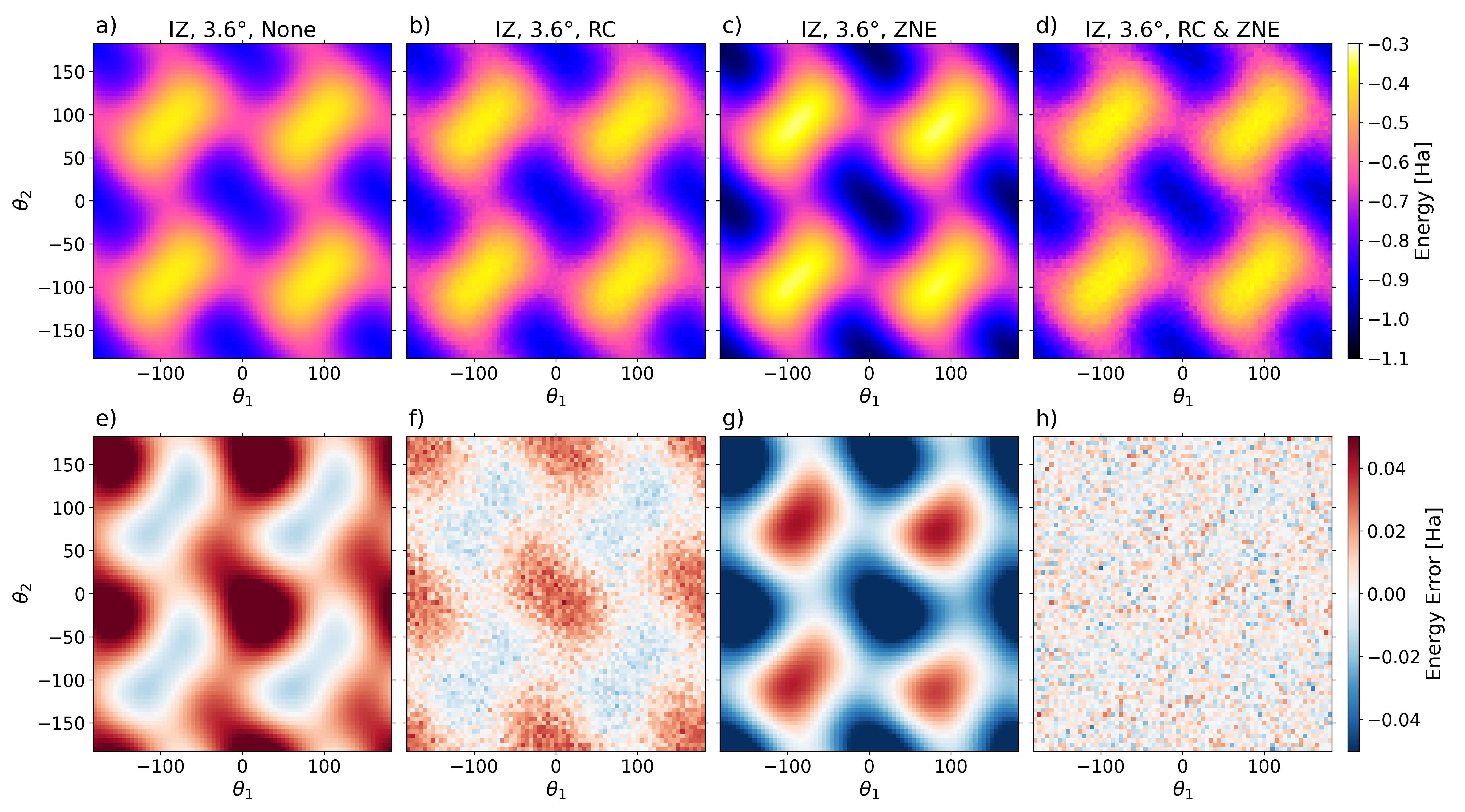}
    \vspace*{-5mm}
    \caption{(a-d) Heatmaps estimated via noisy quantum operation under IZ-rotation noise $(\phi=3.6^\circ)$, (a) without any quantum error mitigation scheme, (b) with only RC, (c) with only ZNE, and (d) with both RC and ZNE. (e-h) Deviation heatmaps from the ideal heatmap in \Cref{fig: heatmap ovrt id insert}(a) for (a-d). The first two parameter $\theta_1, \theta_2$ are vairable while $\theta_0$ is fixed at $8.6^\circ$.}
    \label{fig:heatmap iz}
\end{figure}%

\begin{figure}[!htb]
    \centering
    \includegraphics[scale = 0.104]{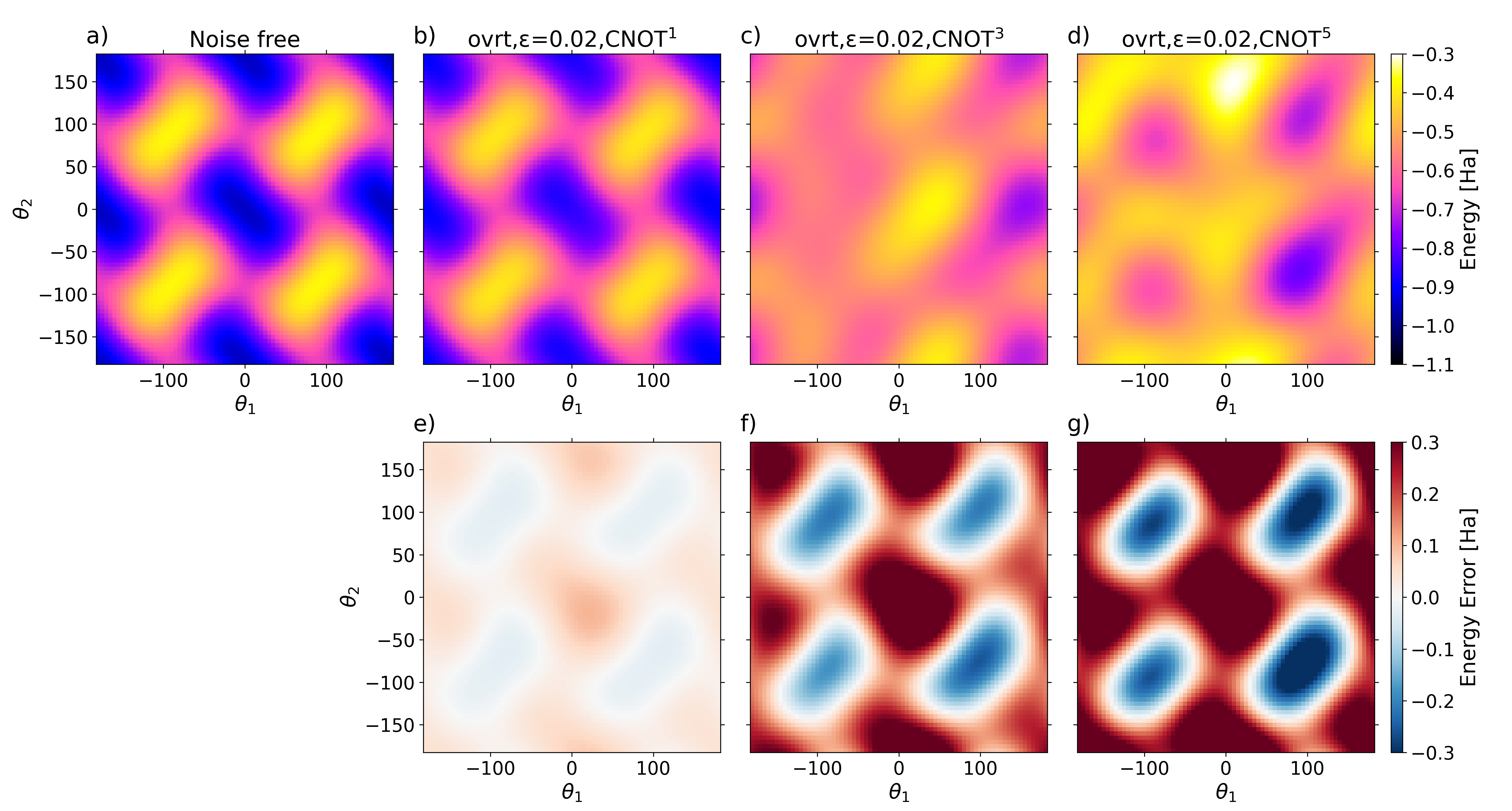}
    \vspace*{-5mm}
    \caption{Heatmaps of the H$_2$ expected energy at different values of the variational parameters. The first two parameter $\theta_1, \theta_2$ are vairable while $\theta_0$ is fixed at $8.6^\circ$, (a) without noise and (b-d) under over-rotation noise $(\epsilon=0.02)$ without quantum error mitigation: (b) original circuit, (c)  noise-magnified version with ($\mathtt{CNOT} \mapsto \mathtt{CNOT}^3$), and (d) noise-magnified version with ($\mathtt{CNOT} \mapsto \mathtt{CNOT}^5$). (e-g) Deviation heatmaps from the ideal heatmap (a) for (b-d).}
    \label{fig: heatmap ovrt id insert}
\end{figure}%

\begin{figure}[!htb]
    \centering
    \includegraphics[scale = 0.104]{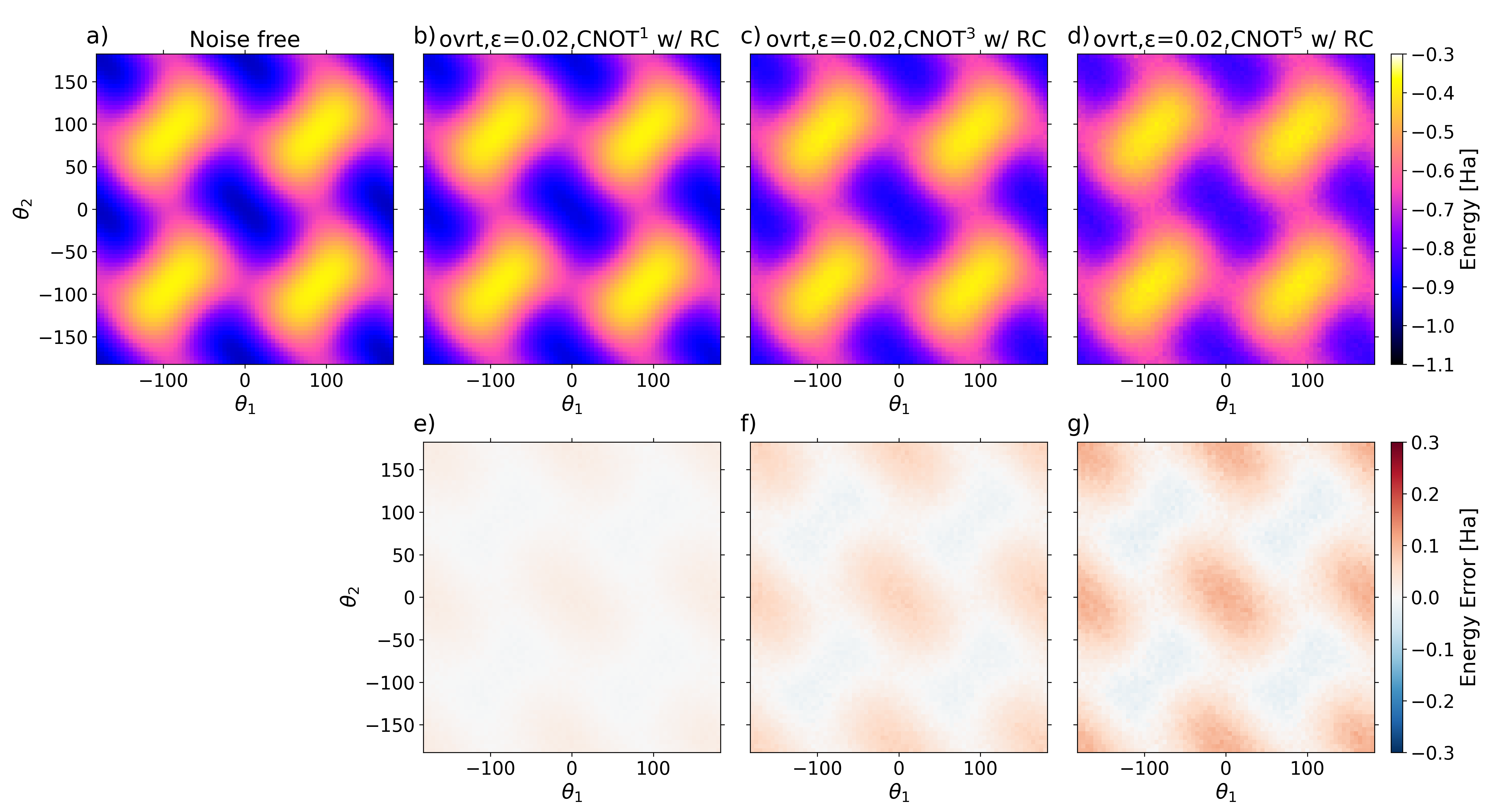}
    \vspace*{-5mm}
    \caption{Heatmaps of the H$_2$ expected energy at different values of the variational parameters. The first two parameter $\theta_1, \theta_2$ are vairable while $\theta_0$ is fixed at $8.6^\circ$, (a) without noise and (b-d) under over-rotation noise $(\epsilon=0.02)$ with RC: (b) original circuit, (c)  noise-magnified version with ($\mathtt{CNOT} \mapsto \mathtt{CNOT}^3$), and (d) noise-magnified version with ($\mathtt{CNOT} \mapsto \mathtt{CNOT}^5$). (e-g) Deviation heatmaps from the ideal heatmap (a) for (b-d).}
    \label{fig: heatmap ovrt id insert w RC}
\end{figure}%

\begin{figure}[h]
    \centering
    \hspace*{-0mm}
    \includegraphics[scale = 0.135]{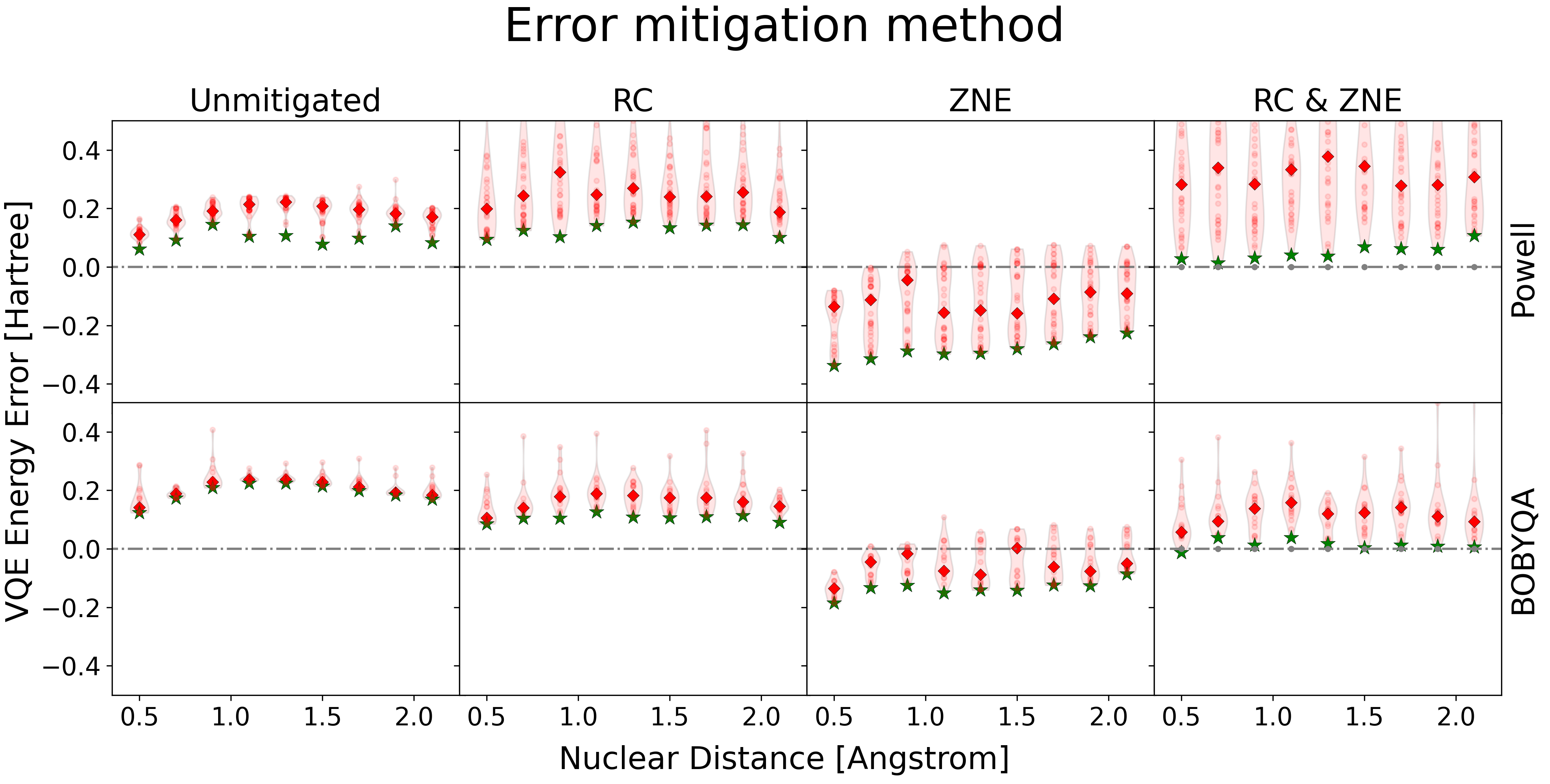}
    \vspace*{-5mm}
    \caption{Comparison between the POWELL and BOBYQA optimizers for simulating the disassociation curve of LiH under the over-rotation noise discussed in the main text. Here randomized compiling is performed with 20 random compilations. Red diamonds (Green stars) indicate the median (minimum) amongst 35 VQE trials with random initial points.}
    \label{LiH_comparison}
\end{figure}

\begin{figure}[h]
    \centering
    \hspace*{-0mm}
    \includegraphics[scale = 0.135]{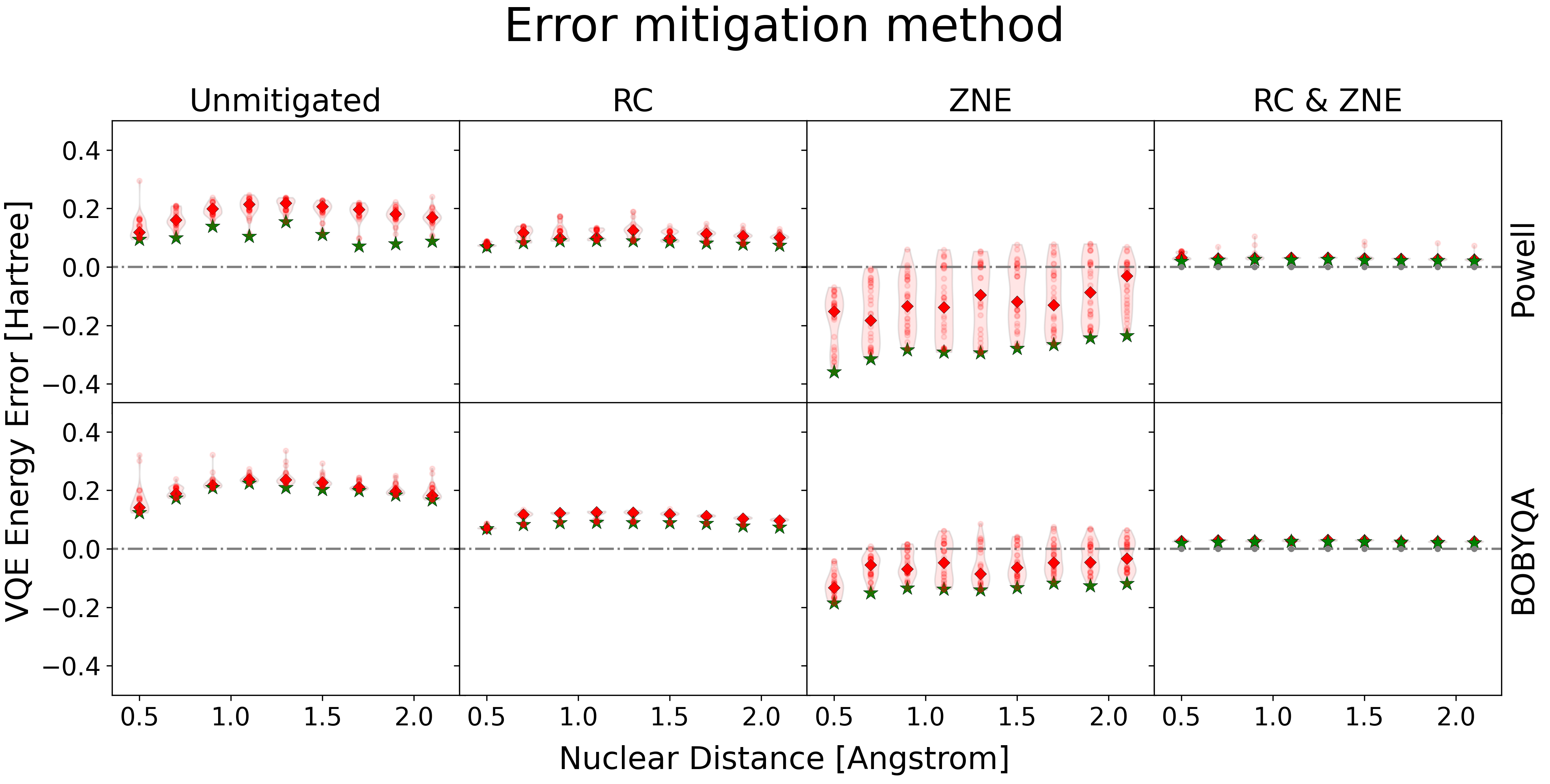}
    \vspace*{-5mm}
    \caption{Comparison between the POWELL and BOBYQA optimizers for simulating the disassociation curve of LiH under over-rotation noise. Here randomized compiling is performed with infinite randomizations by replacing noise sources with their Pauli channel counterparts under twirling \cite{wallman2016noise}. Red diamonds (Green stars) indicate the median (minimum) amongst 35 VQE trials with random initial points.}
    \label{LiH_comparison_inf}
\end{figure}

\begin{figure}[t]
    \centering
    \includegraphics[scale = 0.13]{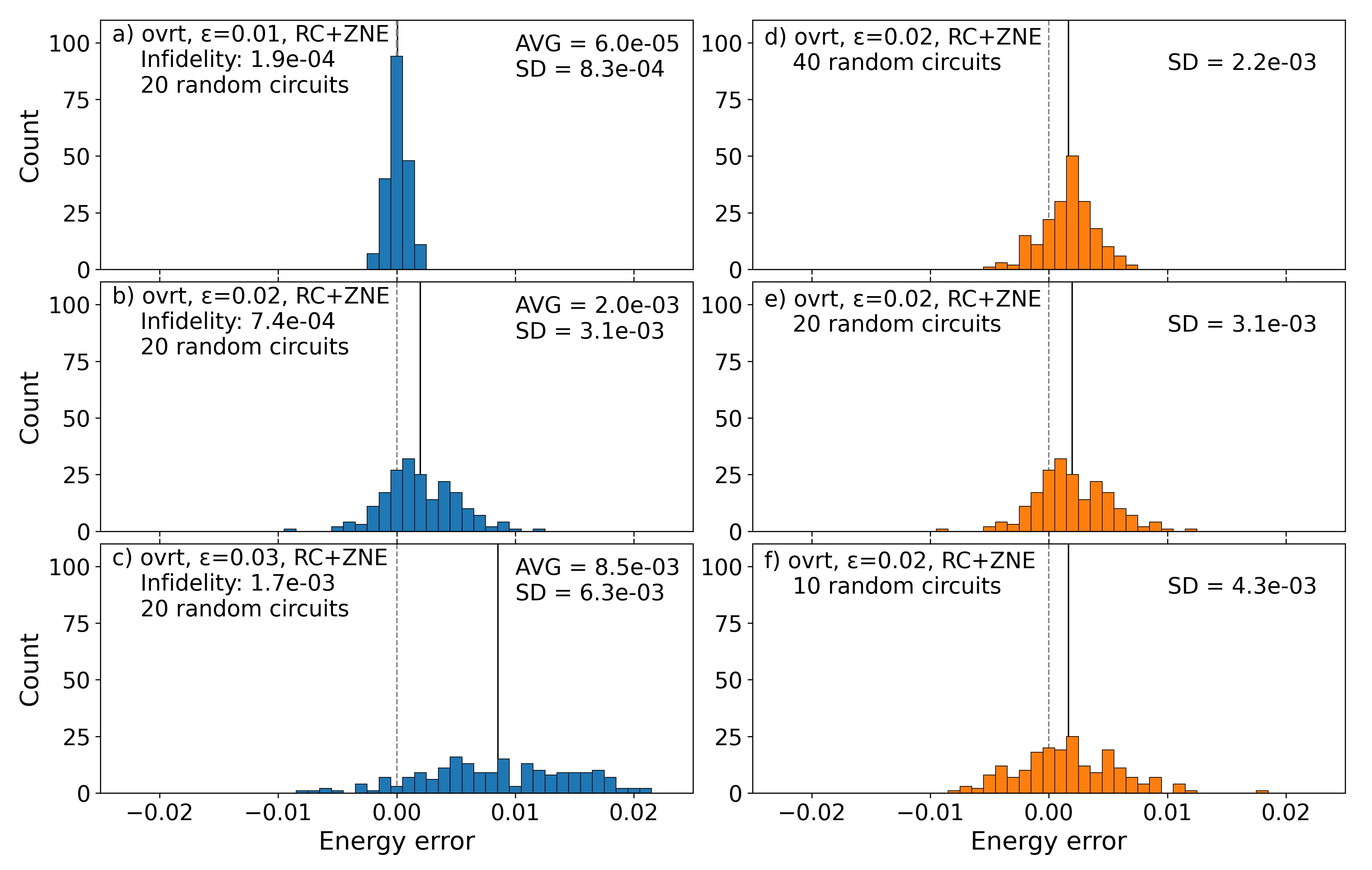}
    \vspace*{-3mm}
    \caption{Histograms of energy errors from ideal values in simulations of H$_2$ with the ansatz circuit with ${\theta_0, \theta_1, \theta_2}={8.6^\circ, 0^\circ ,0^\circ}$. (a-c) Results with various over-rotation amplitudes and fixed number of random circuits for RC, (d-f) Results with various numbers of random circuits for RC and a fixed over-rotation amplitude ($\epsilon=0.02$).}
    \label{precision plot}
\end{figure}%

\begin{figure}[!htb]
    \centering
    \includegraphics[scale = 0.13]{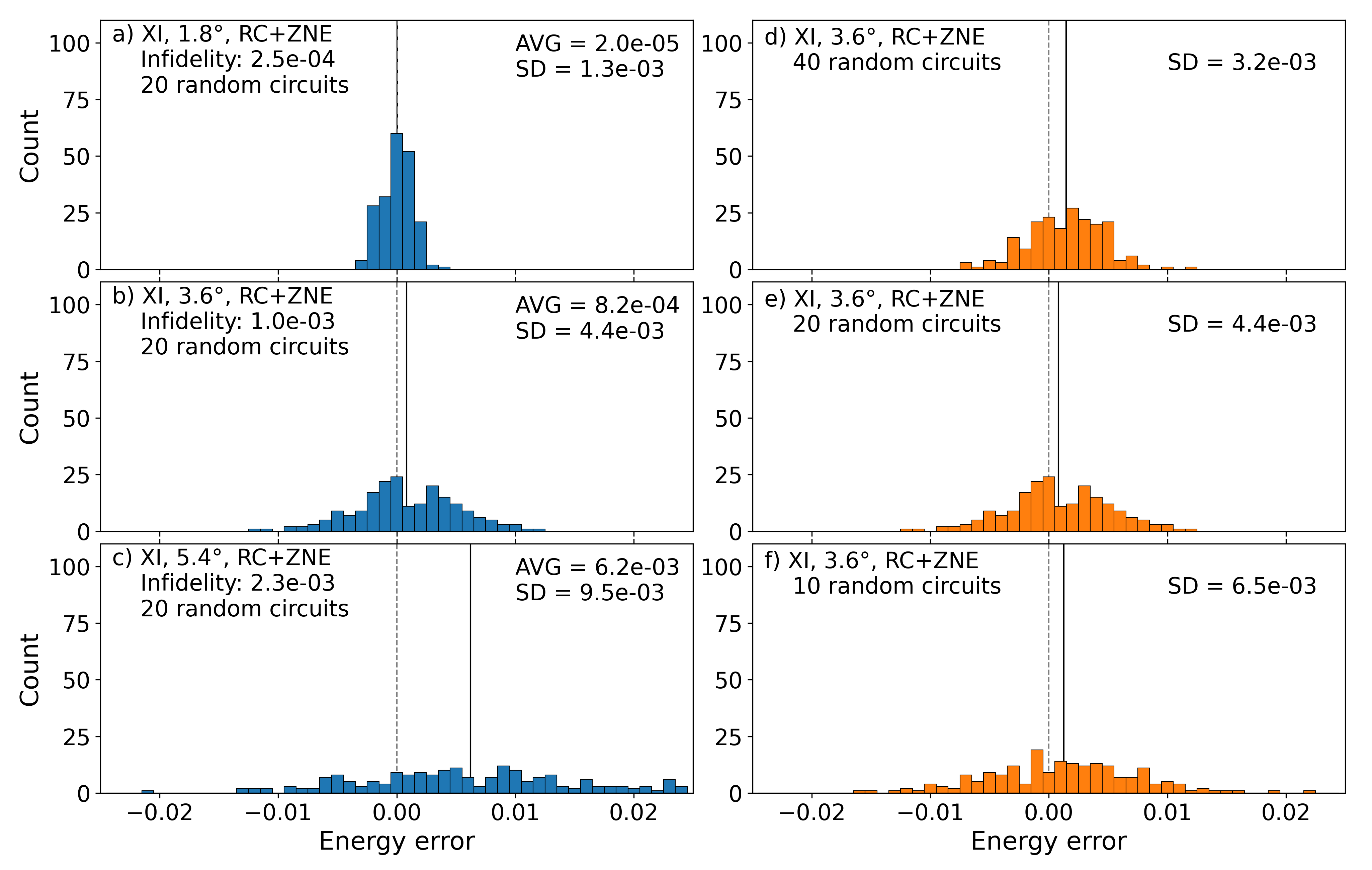}
    \vspace*{-3mm}
    \caption{Histograms of energy errors from ideal values in simulations of H$_2$ with the ansatz circuit with ${\theta_0, \theta_1, \theta_2}={8.6^\circ, 0^\circ ,0^\circ}$. (a-c) Results with various XI-rotation magnitudes $(1.8^\circ,3.6^\circ,5.4^\circ)$ and the fixed number (20) of random circuits for RC, (d-f) Results with various numbers of random circuits for RC and the fixed XI-rotation amplitude ($3.6^\circ$).}
    \label{precision 2}
\end{figure}%

\begin{figure}[!htb]
    \centering
    \includegraphics[scale = 0.13]{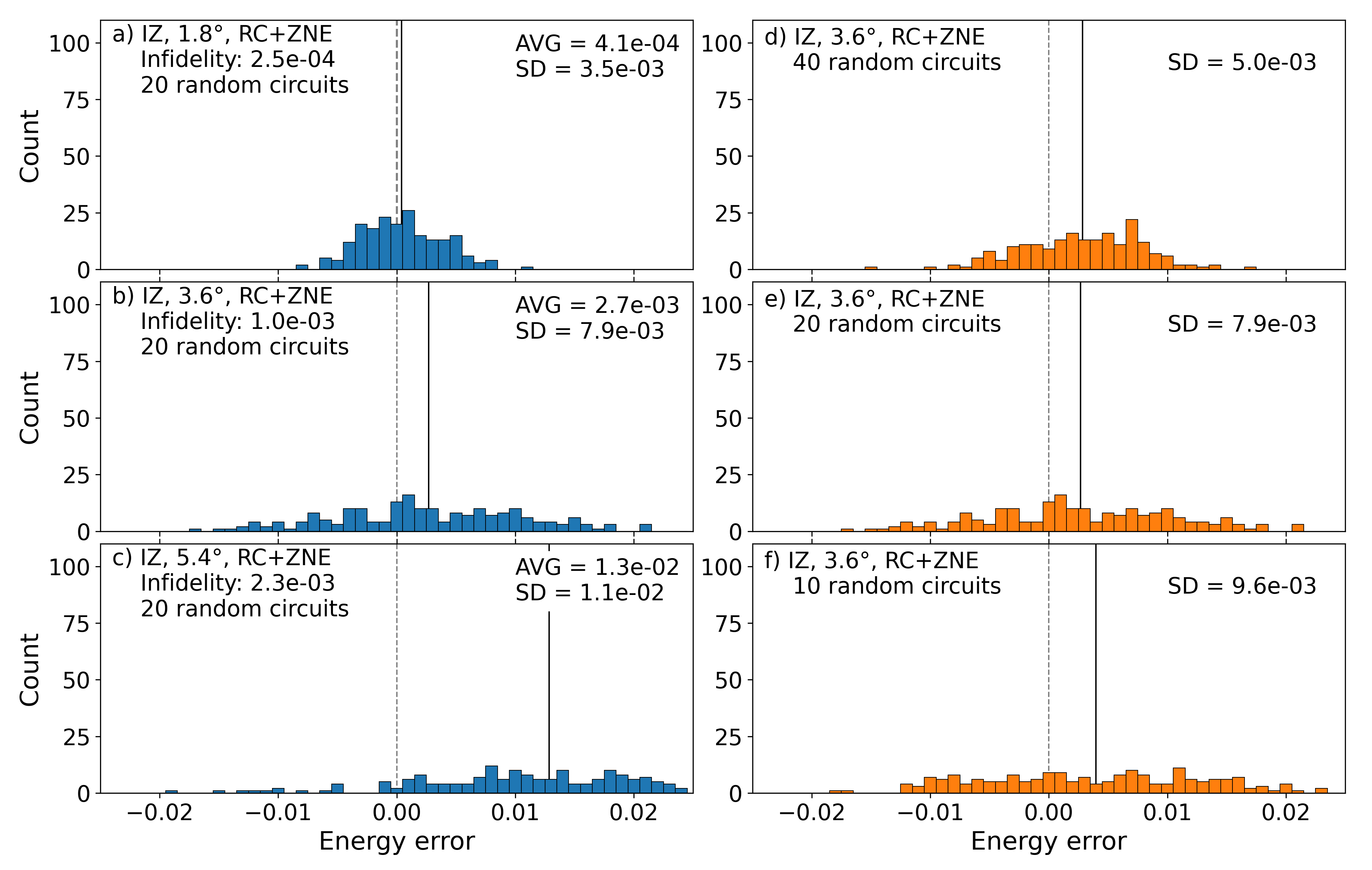}
    \vspace*{-3mm}
    \caption{Histograms of energy errors from ideal values in simulations of H$_2$ with the ansatz circuit with ${\theta_0, \theta_1, \theta_2}={8.6^\circ, 0^\circ ,0^\circ}$. (a-c) Results with various IZ-rotation magnitudes $(1.8^\circ,3.6^\circ,5.4^\circ)$ and the fixed number (20) of random circuits for RC, (d-f) Results with various numbers of random circuits $(10,20,40)$ for RC and the fixed IZ-rotation amplitude (3.6°).}
    \label{precision 3}
\end{figure}%

\begin{figure}[!htb]
    \centering
    \includegraphics[scale = 0.14]{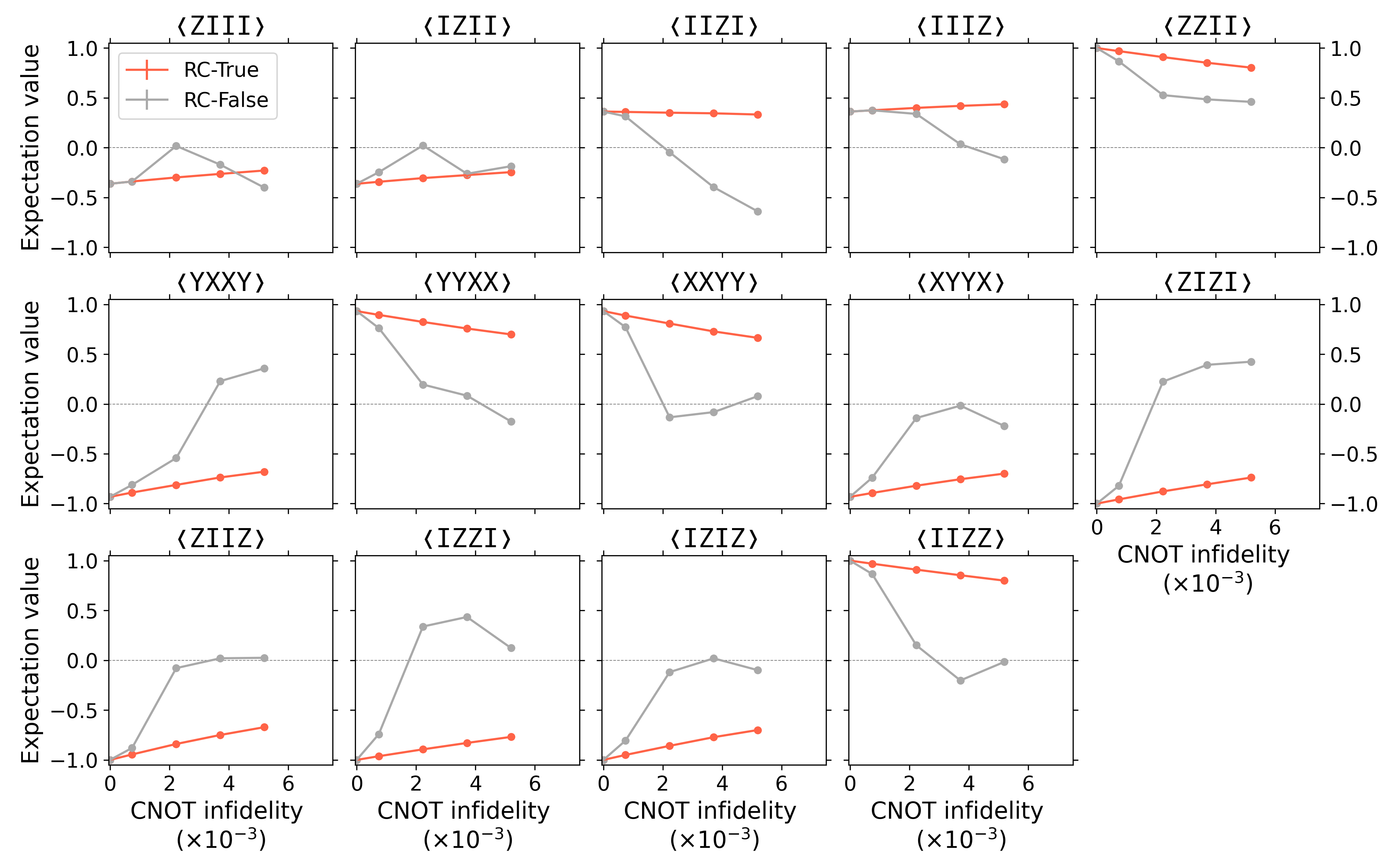}
    \vspace*{-3mm}
    \caption{Expectation values of 14 terms of the H$_2$ Hamiltonian in the final quantum state created by one particular ansatz circuit 
    under over-rotation noise ($\epsilon=0.02$). The $\mathtt{CNOT}$ infidelity is varied by replacing a $\mathtt{CNOT}$ gate with $n$ $\mathtt{CNOT}$ gates for $n \in \{1, 3,5,7\}$.}
    \label{linearized 1}
\end{figure}%

\begin{figure}[!htb]
    \centering
    \includegraphics[scale = 0.14]{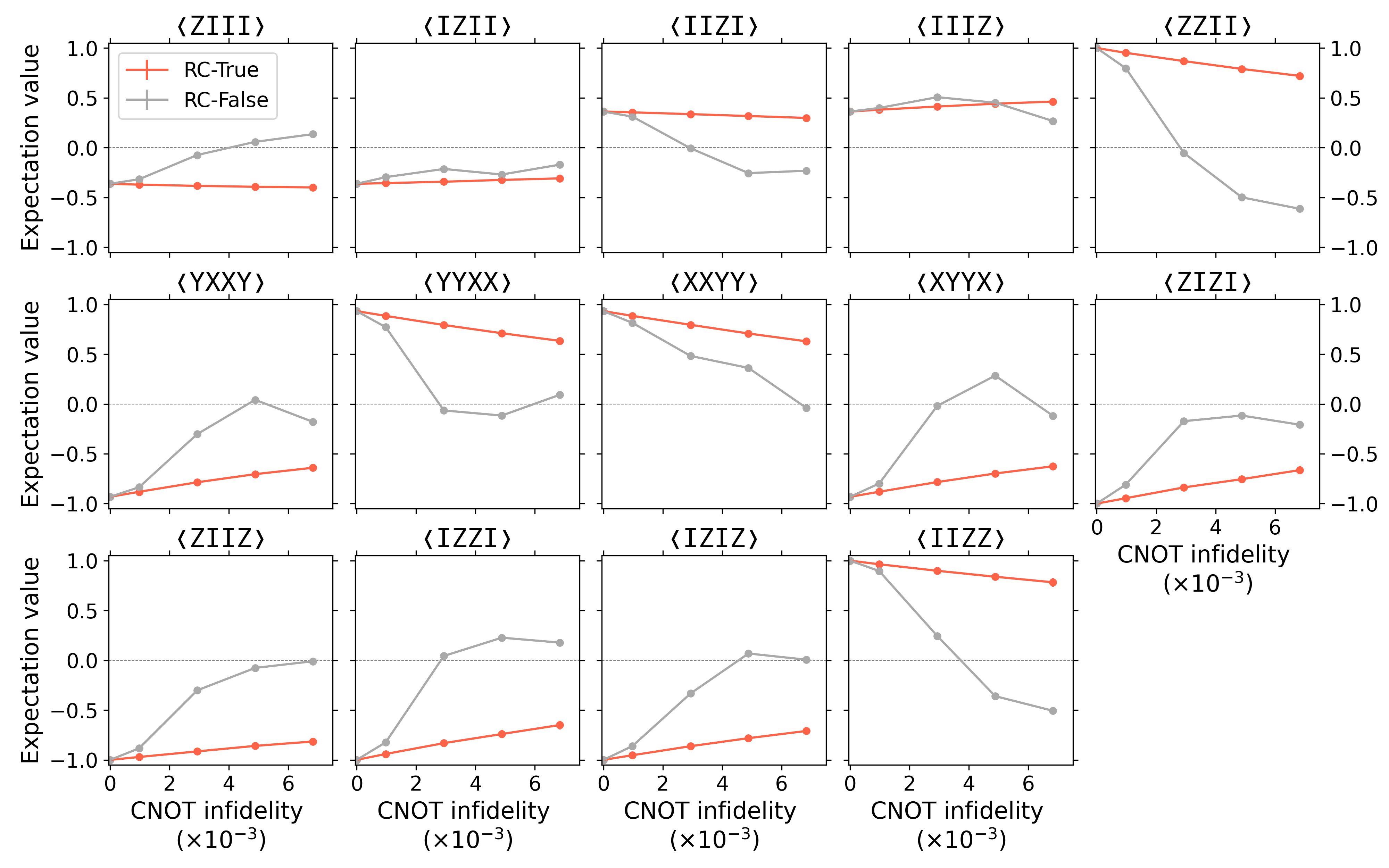}
    \vspace*{-3mm}
    \caption{Expectation values of 14 terms of the H$_2$ Hamiltonian in the final quantum state created by one particular ansatz circuit 
    under XI small-rotation noise ($\theta=3.6^\circ$). The $\mathtt{CNOT}$ infidelity is varied by replacing a $\mathtt{CNOT}$ gate with $n$ $\mathtt{CNOT}$ gates for $n \in \{1, 3,5,7\}$.}
    \label{linearized 2}
\end{figure}%

\begin{figure}[!htb]
    \centering
    \includegraphics[scale = 0.14]{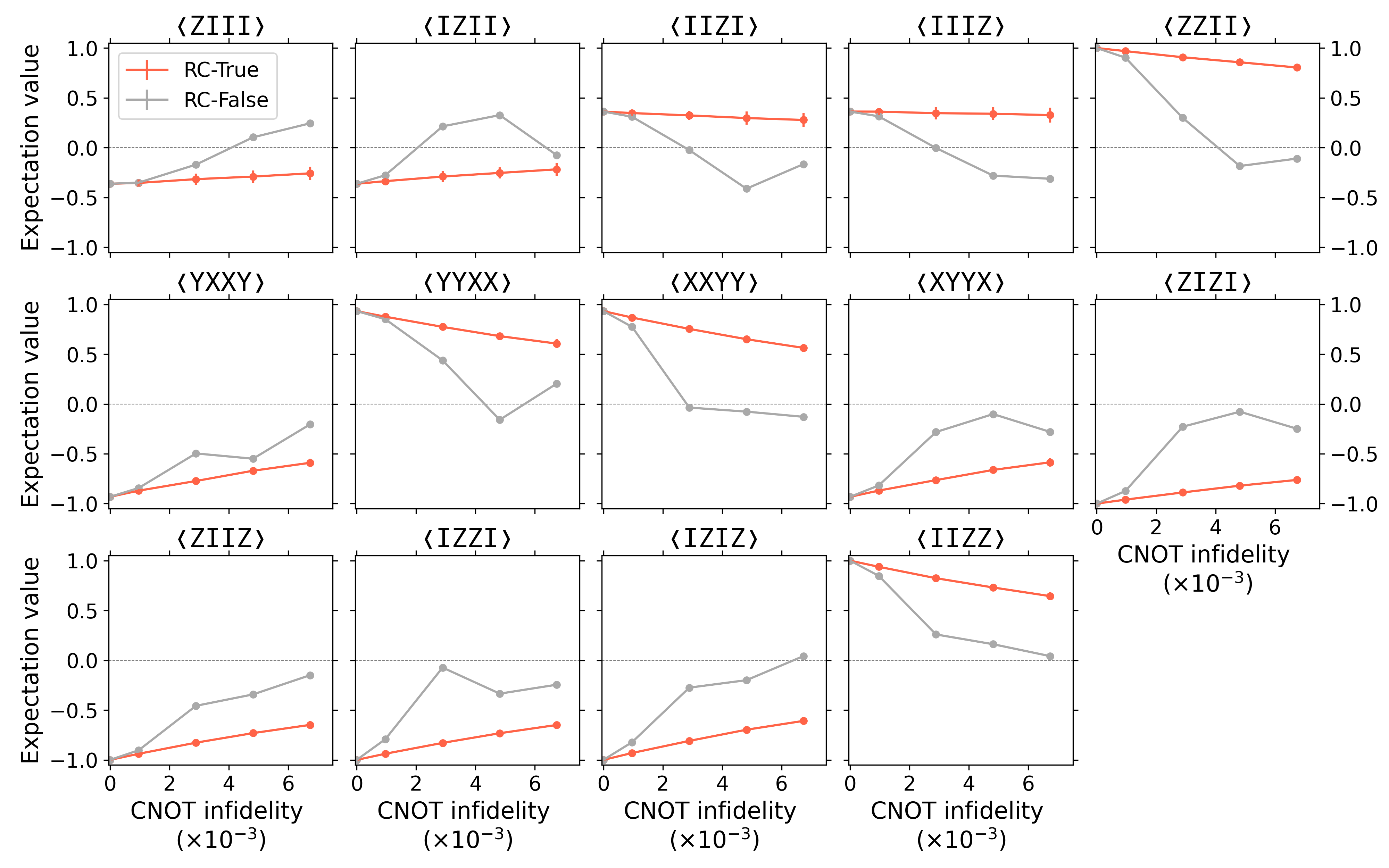}
    \vspace*{-3mm}
    \caption{Expectation values of 14 terms of the H$_2$ Hamiltonian in the final quantum state created by one particular ansatz circuit under IZ small-rotation noise ($\theta=3.6^\circ$). The $\mathtt{CNOT}$ infidelity is varied by replacing a $\mathtt{CNOT}$ gate with $n$ $\mathtt{CNOT}$ gates for $n \in \{1, 3,5,7\}$.}
    \label{linearized 3}
\end{figure}%

\end{document}